\documentclass{aa}  

\usepackage{natbib}
\usepackage{graphicx}
\usepackage{txfonts}
\usepackage{hyperref}
\usepackage{mathrsfs}
\usepackage{mathabx}
\usepackage{color}

\hypersetup{
   colorlinks=true,
   citecolor=blue,
   linkcolor=blue,
   urlcolor=black
   }

\newcommand{\vect}[1]{\textbf{#1}}
\newcommand{\bs}[1]{\boldsymbol{#1}}

\newcommand{\pv}[1]{\widehat{#1}}

\definecolor{emerald}{rgb}{0.3,0.85,0.2}
\definecolor{smcolor}{rgb}{0.7,0.3,0.0}
\newcommand{\smc}[1]{#1}
\newcommand{\jlc}[1]{#1}
\newcommand{\rec}[1]{#1}

\begin{document} 
  \title{Atmospheric thermal tides and planetary spin}
  \subtitle{I. The complex interplay between stratification and rotation}
  \author{P. Auclair-Desrotour\inst{1}
     \and
           S. Mathis\inst{2} \fnmsep \inst{3}
     \and 
           J. Laskar\inst{4}
          }

  \institute{LAB, Université de Bordeaux, CNRS UMR 5804, Université de Bordeaux - Bât. B18N,
              Allée Geoffroy Saint-Hilaire CS50023, 33615 Pessac Cedex\\
              \email{pierre.auclair-desrotour@u-bordeaux.fr}
         \and
             Laboratoire AIM Paris-Saclay, CEA/DRF - CNRS - Université Paris Diderot, IRFU/DAp Centre de Saclay, F-91191 Gif-sur-Yvette Cedex, France\\
             \email{stephane.mathis@cea.fr}
        \and 
              LESIA, Observatoire de Paris, PSL Research University, CNRS, Sorbonne Université, UPMC Univ. Paris 6, Univ. Paris Diderot, Sorbonne Paris Cité, 5 place Jules Janssen, F-92195 Meudon, France
         \and
             IMCCE, Observatoire de Paris, CNRS UMR 8028, PSL Research University, 77 Avenue Denfert-Rochereau, 75014 Paris, France \\
             \email{jacques.laskar@obspm.fr}
             }

  \date{Received ...; accepted ...}

  \abstract
   {Thermal atmospheric tides can torque \smc{telluric} planets away from spin-orbit synchronous rotation, as observed in the case of Venus. They thus participate to determine the possible climates and general circulations of \smc{the atmospheres} of these planets.}
   {The thermal tidal torque exerted on an atmosphere depends both on \smc{its internal structure and rotation,} and the tidal frequency. Particularly, it strongly varies with the \smc{convective} stability of \smc{entropy} stratification. This dependence has to be characterized to constrain \smc{and predict} the \smc{rotational} properties of observed telluric exoplanets. Moreover, it is necessary to validate the approximations \smc{used} in global modelings \smc{such as the traditional approximation, which is used to obtain separable solutions for tidal waves.} }
   {We write the equations governing the dynamics of thermal tides in a local vertically-stratified section of a rotating planetary atmosphere by taking into account the effects of \smc{the complete Coriolis acceleration} on tidal waves. This allows us to derive analytically the tidal torque and the tidally dissipated energy, \rec{which} we use to discuss the possible regimes of tidal dissipation and examine the \smc{key} role played by stratification.}
   {In agreement with early studies, we find that the frequency dependence of the thermal atmospheric tidal torque in the vicinity of synchronization can be approximated by a Maxwell model. This behaviour corresponds to weakly stably stratified or convective fluid layers, as observed in \cite{ADLM2016a}. A strong stable stratification allows gravity waves to propagate, which makes the tidal torque become negligible. \smc{The transition is continuous between these two regimes. The traditional approximation appears to be valid in thin atmospheres and in regimes where the rotation frequency is dominated by the forcing or the buoyancy frequencies.}
    }
 {Depending on the stability of their atmospheres with respect to convection, observed exoplanets can be tidally driven toward synchronous or asynchronous final rotation rates. \smc{The domain of applicability of the traditional approximation is rigorously constrained by calculations.}}


\maketitle


\section{Introduction}

The end of the year 2016 and the beginning of the year 2017 have been marked by the discovery of potentially habitable Earth-like exoplanets orbiting two stars located in the close neighborhood of the Solar system. The first star is Proxima Centauri, where a planet with a minimum mass of $ 1.3 \, M_\Earth $, Proxima~b, has been detected \citep[][]{AE2016,Ribas2016}. The second one is the \smc{ultra-cool} dwarf star Trappist-1, which hosts seven telluric planets of masses between $ 0.2 \, M_\Earth $ and $ 2 \, M_\Earth $, and radii between $ 0.7 \, R_\Earth $ and $ 1.2 \, R_\Earth $ \citep[][]{Gillon2017}. \smc{Proxima b and Trappist-1 e, f, and g} orbit in the habitable zone of their host stars. Besides, they are likely to be covered with an atmosphere like the Earth, Venus or Mars in the Solar system. \smc{Their habitability} is hence strongly constrained by their atmospheric dynamics (\smc{i.e. the} general circulation, zonal winds), which plays a prominent role in the \smc{heat transport} and \rec{determines} \smc{their climate and} surface temperature. 

\smc{The atmosphere dynamics is} tightly related to the rotational dynamics of planets. For instance, the general circulation of rapidly rotating planets such as the Earth is driven by geostrophic flows while planets close to spin-orbit synchronous rotation are subject to large scale convective cells that transport heat from the day side to the night side \citep[e.g.][]{Leconte2013}. It is thus \smc{of great importance} to characterize the rotational evolution of the observed rocky exoplanets. 

Those hosted by Trappist-1 are expected to be either tidally synchronized with the star, or trapped in a higher-order spin-orbit resonance, since they are submitted to a strong tidal gravitational potential \smc{applied on their telluric core} \citep[][]{Gillon2017}. However, as noted by \cite{Ribas2016} in the study of Proxima~b's rotation, \smc{the stellar irradiation can generate thermal atmospheric tides in addition to the standard gravitational tide. Thermal tides} have been proved to be able to compensate the solid tidal torque and to torque the planet \emph{away} from synchronization. This is the case of Venus, which is maintained in a retrograde rotation state of equilibrium by the competition between solid and atmospheric tides \citep[][]{GS1969,DI1980,CL01,CL2003,ADLM2016b}. 

In a previous study, by using a global ab initio modeling, we investigated the role played by the structure and properties of the atmosphere in response to a semidiurnal quadrupolar tidal forcing \citep[][]{ADLM2016a}. Particularly, we showed that stable stratification \smc{is} able to modify the dependence of the atmospheric tidal torque on the tidal frequency. These results suggest that the Maxwell-like\footnote{So called in reference to the Maxwell model \citep[see e.g.][]{Greenberg2009,Correia2014}.} frequency dependence obtained by early works with \smc{parametrized} models \citep[][]{ID1978,CL01} or General Circulation Models \citep[or GCMs,][]{Leconte2015} corresponds to the tidal response of \rec{a} neutrally-stratified atmosphere with respect to convection. \smc{They also show} that a strong stable stratification tends to annihilate the global tidal bulge by allowing gravity waves to propagate, so that the resulting \smc{atmospheric} tidal torque becomes negligible. This has \jlc{large} repercussions on the final rotation rate of the planet. In the neutrally-stratified case, an asynchronous rotation rate can be reached, while the solid tide dominates in the second case, leading the body to spin-orbit synchronization.

However, these tendencies were obtained by considering asymptotic cases (neutrally \smc{slowly rotating} and isothermal stably stratified atmospheres), which let the details of \smc{the continuous variation} of the tidal response \smc{with} stratification aside. Moreover, the global model used to describe the dynamics of tidal waves in this early work is based upon an approximation on the hierarchy between rotation and stratification, the \emph{traditional approximation}. \smc{It consists in neglecting the latitudinal projection of the rotation vector in the Coriolis acceleration}, \smc{which allows to integrate separately the vertical and horizontal structure of the tidal response.} The traditional approximation is not valid for \jlc{all} configurations and its domain of applicability is not clearly established \citep[][]{GZ2008,Tort2014}. Some of \jlc{the} limitations \jlc{of the traditional approximation} have been identified in geophysical and astrophysical fluid dynamics \citep[][]{GS2005}. Particularly, \jlc{the general case is fully 2D and thus} the coordinates cannot be separated \citep[][]{MNT2014}. \smc{Hence, the traditional approximation} needs to be validated \smc{for thermal atmospheric tides}. \smc{It} is \smc{thus} necessary to develop a tractable model enabling us to explore widely the \smc{possible} stratifications and rotations without any approximation on the Coriolis acceleration. \smc{Such a model would also allow us to} \rec{discuss} the different tidal regimes, their consequences on the evolution of the rotation rate of a planet, and the \smc{possible} bias \rec{resulting from} approximations made in the global modeling. 

\smc{Therefore}, we propose in this study to characterize the behaviour of the atmospheric tidal torque with respect to stratification in the framework of a \smc{2D} simplified local Cartesian modeling where the effect of rotation is fully taken into account. This approach is based on the early works by \cite{GS2005}, \rec{ \cite{MNT2014} and \cite{Andre2017}}, who examine the effect of stratification and rotation on waves propagating within planetary \smc{oceans}, stars \rec{and gaseous envelopes of giant planets}. \smc{We generalize their formalism to treat the case of stratified atmospheres with vertically-dependent hydrostatic equilibrium structures.}

We introduce the physical setup of the local model in Section~\ref{sec:physical_setup} and write the equations describing the dynamics of thermally generated tidal waves in a planetary fluid layer in Section~\ref{sec:dynamics}. \smc{Then} we derive, in Section~\ref{sec:energy_balance}, the energy balance associated to a propagating tidal mode. In Section~\ref{sec:analytic_solutions}, analytic solutions are computed in the case of \smc{simplified atmospheric models, namely} a homogeneous fluid with uniform background distributions and of an isothermal stably-stratified atmosphere. In Section~\ref{sec:tidal_regimes}, these solutions are used to discuss the possible tidal regimes and the dependence of the \smc{atmospheric} tidal torque on stratification \smc{in the non-traditional framework}. We show that beyond a critical value of the Brunt-Väisälä frequency this torque becomes negligible. \smc{We then} illustrate the consequences of this behaviour on the planet's rotation by computing the evolution of this later for various stratifications. We end this work with a study of the applicability of the traditional approximation in Section~\ref{sec:app_trad} \smc{that} allows us to identify asymptotic regimes where it \smc{can be applied}. \smc{Finally, we} give our conclusions in Section~\ref{sec:conclusions}.

  
\section{Physical setup and background structure}
\label{sec:physical_setup}

\begin{figure}
   \centering
   \includegraphics[width=0.45\textwidth]{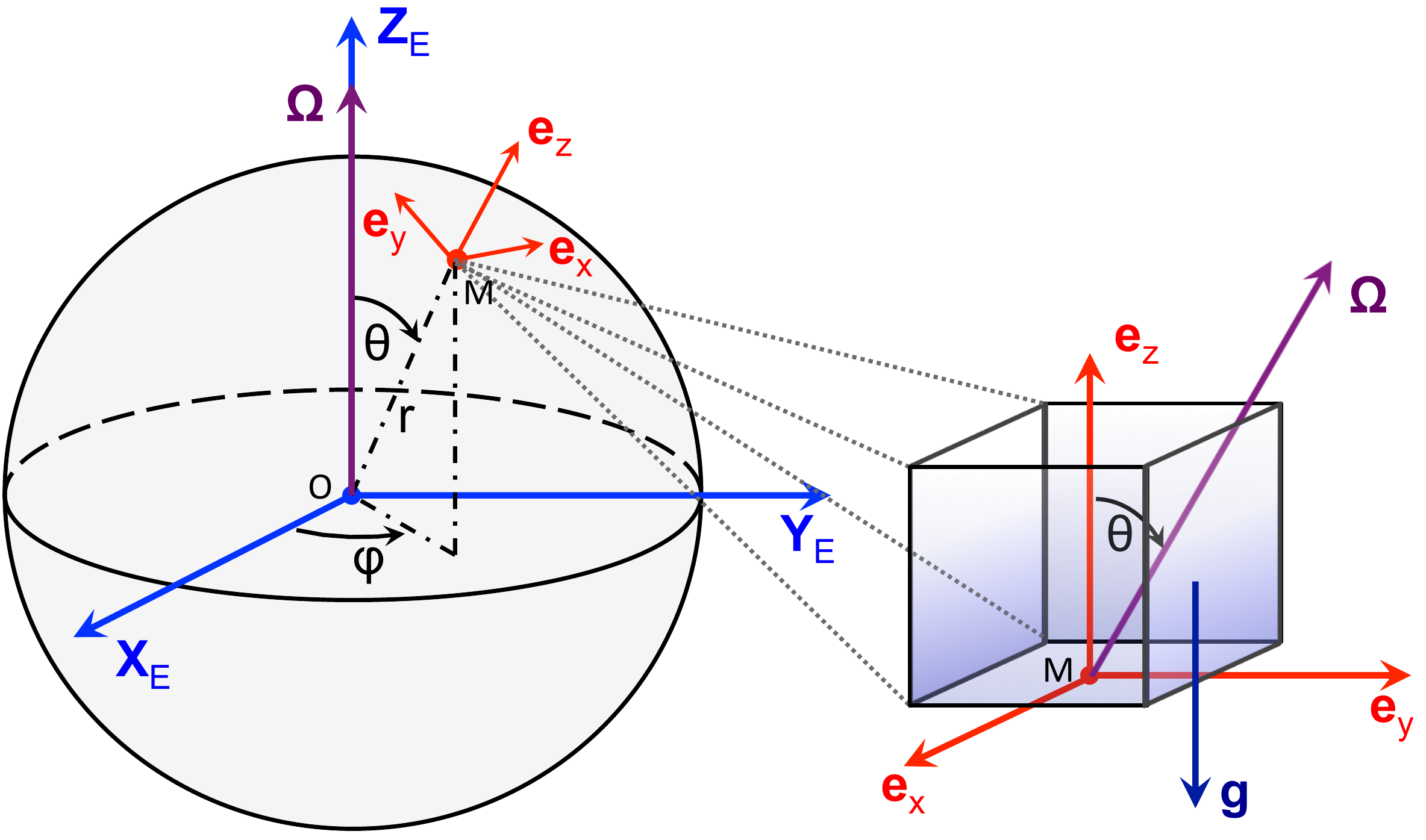}
   \caption{Spherical and Cartesian reference frames and systems of coordinates. \smc{The vectors $ \boldsymbol{\Omega} $ and $ \textbf{g} $ designate the rotation and the gravity respectively}.}
              \label{fig:box}%
    \end{figure}

Following \cite{GS2005} and \cite{MNT2014}, we consider a local volume within a planetary \smc{atmospheric} layer (Fig.~\ref{fig:box}). This volume is a Cartesian box of side $ L $ such that $ L \ll R $, $ R $ being the \smc{planet radius}. \smc{It} is rotating uniformly at the angular velocity $ \Omega $. We \rec{denote by} $ \boldsymbol{\Omega} $ the corresponding spin vector. The position of the fluid box is located in the co-rotating frame attached to the body $ \mathcal{R}_{\rm E} : \left\{ O, \vect{X}_{\rm E} , \vect{Y}_{\rm E}, \vect{Z}_{\rm E} \right\} $, such that $ \vect{Z}_{\rm E} = \bs{\Omega}/\left| \Omega \right| $, with the usual system of spherical coordinates $ \left( r , \theta , \varphi \right) $, where $ r $ stands for the \smc{vertical coordinate}, $ \theta $ for the colatitude and $ \varphi $ for the longitude. The associated vectorial basis is denoted $ \left( \vect{e}_r , \vect{e}_\theta , \vect{e}_\varphi \right) $. To study the dynamics of tidal waves inside the box, we use the Cartesian coordinates $x $ (west-east), $ y $ (south-north) and $ z $ (vertical, positive upward), and the associated basis $ \left( \vect{e}_x , \vect{e}_y , \vect{e}_z \right) = \left( \vect{e}_\varphi , - \vect{e}_\theta , \vect{e}_r \right) $. \rec{Finally}, the time is denoted $ t$. For convenience, we will use as much as possible the same notations as \cite{GS2005} for physical quantities.

The structure of the layer is described by the background spatial distributions of gravity $ g $, pressure $ p_0 $, density $ \rho_0 $ and \rec{temperature} $ T_0 $. To simplify \smc{the problem}, we assume that these quantities vary with the \smc{vertical} coordinate \smc{($z $)} only. This \smc{\rec{corresponds to ignoring}} the effects of centrifugal distortion and the day-night variations of the \smc{atmosphere}. We also introduce the pressure \rec{scale height} of the fluid,

\begin{equation}
H \left( z \right) = \frac{p_0}{g \rho_0},
\end{equation}

\noindent and its Brunt-Väisälä frequency, i.e. the frequency characterizing the stability of the vertical \smc{entropy} stratification, defined as 

\begin{equation}
N^2 \left( z \right) = g \left[ \frac{1}{\Gamma_1} \frac{d \ln p_0}{dz} - \frac{d \ln \rho_0}{dz} \right],
\end{equation}

\noindent where $ \Gamma_1 $ stands for the first adiabatic exponent \citep[e.g.][]{GZ2008}. We will consider in the following that the fluid is a perfect \rec{gas}, so that $ \Gamma_1 = 1.4 $. As the \smc{atmosphere} is supposed to be in solid rotation with the body, mean flows are ignored in this \smc{first paper}. However, the internal dissipation due to radiative cooling is taken into account. \smc{In the Maxwell approximation mentioned above, the maximal amplitude of the atmospheric torque corresponds to \jlc{equalizing the tidal period $ \tau_{\rm tide} $ and an effective thermal time} associated with internal diffusive and radiative processes $ \tau_0 $ \citep[][]{Leconte2015,ADLM2016a}. These processes can be described at first order by a \emph{Newtonian cooling} as in \cite{ADLM2016a} \smc{\citep[see also][]{LM1967,Dickinson1968}}.} This introduces a parameter for the efficiency of dissipation, the frequency $ \sigma_0 $, which is the \smc{inverse} of the local radiative time of the fluid. Like the other quantities, $ \sigma_0 $ can vary along the \smc{vertical} direction \citep[see for example][for Venus]{PY1975}\smc{; it} will be set to a constant \jlc{value} here for the sake of simplicity. The effect of a radiative time varying with altitude on the tidal perturbation has been studied in the framework of the classical theory of atmospheric tides \citep[e.g.][]{Dickinson1968,LM1967}.

\section{\rec{Tidal wave dynamics}}
\label{sec:dynamics}

The fluid is forced thermally by the perturber with the thermal power per unit mass $ J $. The resulting perturbed quantities are the variations of pressure $ \delta p $, density $ \delta \rho $ and the velocity field $ \textbf{u} = \left( u , v , w \right) $. To simplify the equations of dynamics, we will use the \smc{reduced} pressure variations $ p = \delta p / \rho_0 $ and the buoyancy $ b = - g \delta \rho / \rho_0 $ instead of $ \delta p $ and $ \delta \rho $ \citep[see e.g.][]{GS2005}. The effect of rotation on the tidal perturbation is taken into account through the Coriolis acceleration. Hence, we introduce the Coriolis parameters $ f = 2 \Omega \cos \theta $ and $ \tilde{f} = 2 \Omega \sin \theta $ and suppose them to be constant in the box, which is the so-called \emph{f-plane approximation}. Note that \smc{$ f $ corresponds to the vertical projection of the rotation vector and $ \tilde{f} $ to its \smc{latitudinal} component, which is} responsible for the coupling of the \smc{latitudinal} and vertical structures of tidal waves. To make the problem \smc{separable and} tractable analytically \smc{within} spherical geometry, it is necessary to assume the \emph{traditional approximation}, i.e. to ignore this term. Here, the simplified Cartesian geometry of the local modeling allows us to conserve \smc{the complete Coriolis acceleration that includes terms in $\tilde{f}$}.

\smc{We assume that asynchronous rotation rates are possible only in the vicinity of synchronous rotation, in range of forcing periods defined by $ \tau_{\rm tide} \sim \tau_0 $. This allows us to ignore the effects of compressibility and to simplify calculations by applying the \emph{anelastic approximation} \citep[][]{SV1960}. Thus, following \cite{MNT2014}, we neglect the contribution of acoustic waves. Note that compressibility can nevertheless contribute to the atmospheric tidal response in a non-negligible way in the regime of rapidly rotating bodies, where horizontally propagating acoustic waves can be generated, the so-called \emph{Lamb modes}}. Finally, we assume the so-called \emph{Cowling approximation} \citep[][]{Cowling1941}, which consists in ignoring the perturbation of the self-gravitational potential. The dynamics of the tidally forced fluid response is thus described by the \smc{following} linearized Navier-Stokes equation

\begin{equation}
 \textbf{u}_t + 2 \boldsymbol{\Omega} \times \textbf{u} = - \nabla p + b \, \textbf{e}_z,
 \label{prim_momentum}
\end{equation}

\noindent the equation of mass conservation

\begin{equation}
\nabla \cdot \left( \rho_0 \textbf{u} \right) = 0,
\label{prim_mass}
\end{equation}

\noindent and the equation of energy

\begin{equation}
 b_t  + N^2 w = \frac{\kappa}{H} J - \sigma_0 b,
\label{prim_buoyancy}
\end{equation}

\noindent where the subscript $ t $ denotes the \rec{partial derivative} in time, $ \kappa = \left( \Gamma_1 - 1 \right) / \Gamma_1 $, and $ \sigma_0 b $ stands for the sink term associated with the Newtonian cooling. Because of the periodicity in time of the tidal forcing \smc{(thermal and gravitational)}, any perturbed quantity $ q $ can be expanded as a Fourier series of the form

\begin{equation}
q \left( \textbf{x} , t \right) = \sum_{\sigma} q^\sigma \left( \textbf{x} \right) e^{i \sigma t},
\label{Fourier}
\end{equation}

\noindent the parameter $ \sigma $ being the tidal frequency of a component and $ q^\sigma $ the Fourier coefficient of the expansion. In the following, the superscript $ \sigma $ will be omitted in order to lighten expressions. By substituting Eq.~(\ref{Fourier}) in Eqs.~(\ref{prim_momentum}-\ref{prim_buoyancy}), we end up with the system of linearized primitive equations

\begin{eqnarray}
\label{eqprim1}
i \sigma u - f v + \tilde{f} w & = & - p_x , \\
i \sigma v + f u & = & - p_y ,  \\
\label{eqprim3}
i \sigma w - \tilde{f} u & = & -p_z + b , \\
u_x + v_y + w_z + \frac{d \ln \rho_0}{dz} w & = & 0, \\
\label{eqprim5}
 \left( i \sigma + \sigma_0 \right) b + N^2 w  & = & \frac{\kappa}{H} J. 
\end{eqnarray}

\noindent Note that the notations $ u $, $ v $, $ w $, $p $ and $b$ now refer to the spatial distributions of perturbed quantities.

We consider waves propagating in the horizontal direction \smc{along the} vector $ \textbf{e}_\alpha = \cos \alpha \, \textbf{e}_x + \sin \alpha \, \textbf{e}_y$, $ \alpha $ being the angle of the direction \smc{of propagation in the $ \left( \textbf{e}_x , \textbf{e}_y \right) $ plane}. Hence, using the change of variable $ \chi = x \cos \alpha  + y \sin \alpha $, the system of Eqs.~(\ref{eqprim1}-\ref{eqprim5}) can be reduced to a single equation \smc{for} $ w $,

\begin{equation}
A w_{\chi \chi} + 2 B w_{\chi z} + C w_{zz} + D w_{\chi} + E w_{z} + F w = S,
\end{equation}

\noindent where the subscripts $ \chi $ and $ z $ refer to the \smc{horizontal and vertical derivatives respectively}. The coefficients associated with second-order derivatives are expressed as

\begin{align}
& A \left( z \right) = N^2 + \eta^{-1} \left( f_{\rm s}^2 - \sigma^2 \right), \\
& B \left( z \right) = \eta^{-1} f  f_{\rm s}, \\
& C \left( z \right) = - \eta^{-1} \left( \sigma^2 - f^2 \right),
\end{align}

\noindent those associated with first order derivatives as

\begin{align}
& D \left( z \right) = \eta^{-1} \frac{d \ln \rho_0}{dz} \tilde{f} \left( i \sigma \cos \alpha + f \sin \alpha \right), \\
& E \left( z \right) = \eta^{-1}  \left( f^2 - \sigma^2 \right) \frac{d \ln \rho_0}{dz}, \\
& F \left( z \right) = \eta^{-1} \left( f^2 - \sigma^2 \right) \frac{d}{dz} \left( \frac{d \ln \rho_0}{dz} \right),
\end{align}

\noindent and the source term $ S $ due to the thermal forcing \rec{is written} 

\begin{equation}
S \left( z , \chi \right) = \frac{\kappa}{H} J_{\chi \chi}. 
\end{equation}

\noindent In these expressions, we \smc{have} \rec{introduced} the modified Coriolis parameter \rec{of} \cite{GS2005}, $ f_{\rm s} = \tilde{f} \sin \alpha $, and the \smc{function}

\begin{equation}
\eta \left( \sigma \right) = \frac{\sigma}{\sigma - i \sigma_0},
\end{equation}

\noindent which is such that $ \eta = 1 $ if the radiative cooling is ignored and $ 0 \leq \left| \eta \right| < 1 $ otherwise. In the asymptotic regime dominated by the radiative cooling ($ \left| \sigma / \sigma_0 \right| \rightarrow 0 $), $ \eta \rightarrow 0 $. In the regime where thermal time is far greater than the tidal period ($ \left| \sigma / \sigma_0 \right| \rightarrow + \infty  $), $ \eta \rightarrow 1 $. 

Hence, by setting $ \sigma_0 = 0 $ and a uniform density profile ($ d \ln \rho_0 / dz = 0 $), we eliminate terms associated with first-order derivatives and recover the equations given by \cite{GS2005}. 

Following this early work, we seek \smc{2D} solutions expressed as

\begin{equation}
w \left( z , \chi \right) = \Psi \left( z \right) e^{i \left[ k_\perp \chi + \delta \left( z \right) \right]},
\end{equation}

\noindent where \smc{$ k_\perp $} stands for the horizontal wavenumber, $ \delta $ is the function defined as 

\begin{equation}
\delta \left( z \right) = k_\perp \frac{f f_{\rm s}}{\sigma^2 - f^2} z + i \frac{1}{2} \ln \rho_0,
\end{equation}

\noindent and $ \Psi $ the solution of the Schrödinger-like vertical structure equation 

\begin{equation}
\frac{d^2 \Psi}{dz^2} + k_z^2 \Psi = - k_\perp^2 \frac{\pv{S}}{C} \Phi^{-1}.
\label{vert_struct}
\end{equation}

In this equation, \rec{$ \pv{S} $ denotes the vertical profile of $ S $ resulting from the above separation of coordinates and} $ k_z $ the \rec{local vertical wavenumber} of the mode, defined by

\begin{equation}
\begin{array}{lcl}
k_z^2  & = & \displaystyle k_\perp^2 \left[ \frac{\eta N^2 - \sigma^2}{\sigma^2 - f^2} + \left( \frac{\sigma f_{\rm s}}{\sigma^2 - f^2} \right)^2  \right] \\[0.3cm]
 & & \displaystyle + \frac{d \ln \rho_0}{dz} \left( k_\perp \frac{\sigma \tilde{f} \cos \alpha}{\sigma^2 - f^2} - \frac{1}{4} \frac{d \ln \rho_0}{dz} \right) + \frac{1}{2} \frac{d^2 \ln \rho_0}{dz^2},
\end{array}
\label{kz2}
\end{equation}

\noindent \smc{while} $ \Phi \left (z \right) = \exp \left( i \delta \right) $. We recognize in the first term of Eq.~(\ref{kz2}) the classical \smc{vertical} wavenumber of gravito-inertial waves. Other terms are associated with the variation of background distributions. The polarization relations giving the vertical profiles of perturbed quantities as functions of $ \Psi $ and its first derivative are deduced straightforwardly from primitive equations. \rec{Denoting by $ \pv{q} $ the vertical profile of a quantity $q $ (such that $ q \left( z , \chi \right) = \pv{q} \left( z \right) e^{i k_\perp \chi} $),} we thus obtain, for the buoyancy,

\begin{equation}
\pv{b} \left( z \right) = - \frac{i}{\sigma - i \sigma_0} \left[ \frac{\kappa}{H} \pv{J} - N^2 \Phi \Psi \right],
\label{bz}
\end{equation}

\noindent for the pressure,

\begin{equation}
\pv{p} \left( z \right) = i \frac{f^2 - \sigma^2}{\sigma k_\perp^2} \Phi \left[ \frac{d \Psi}{dz} + \left( \frac{1}{2} \frac{d \ln \rho_0}{dz} + \mathcal{A} \right) \Psi \right],
\label{pz}
\end{equation}

\noindent and for the components of the velocity field,

\begin{align}
\label{uz}
& \pv{u} \left( z \right) = \frac{i \sigma \cos \alpha + f \sin \alpha}{\sigma k_\perp} \Phi \left[  \frac{d \Psi}{dz} + \left( \frac{1}{2} \frac{d \ln \rho_0}{dz} + \mathcal{B} \right) \Psi \right], \\[0.3cm]
\label{vz}
& \pv{v} \left( z \right) = - \frac{f \cos \alpha - i \sigma \sin \alpha}{\sigma k_\perp} \Phi \left[ \frac{d \Psi}{dz} + \left( \frac{1}{2} \frac{d \ln \rho_0}{dz} + \mathcal{C} \right) \Psi \right], \\[0.3cm]
\label{wz}
& \pv{w} \left( z \right) = \Phi \Psi,
\end{align}

\noindent with the coefficients

\begin{align}
& \mathcal{A} =  - k_\perp \frac{\sigma \tilde{f} \cos \alpha}{\sigma^2 - f^2}, \\
& \mathcal{B} =   i k_\perp f_{\rm s} \left[  \frac{f}{\sigma^2 - f^2} + \frac{\sin \alpha}{i \sigma \cos \alpha + f \sin \alpha} \right], \\
& \mathcal{C} =  i k_\perp f_{\rm s} \left[ \frac{f}{\sigma^2 - f^2} + \frac{\cos \alpha}{f \cos \alpha - i \sigma \sin \alpha} \right].
\end{align}

\section{Energy balance}
\label{sec:energy_balance}

The vertical profiles of the perturbed quantities \smc{being} established \smc{in Eqs.~(\ref{bz}-\ref{wz})}, we can compute the energy balance \smc{and torque} associated with the \smc{atmospheric tide}. The equation \smc{for} the energy is obtained by multiplying the momentum equation \smc{Eq.}~(\ref{prim_momentum}) by $ \textbf{u} $ and the equation of buoyancy \smc{Eq.}~(\ref{prim_buoyancy}) by $ b $. \smc{We get:}

\begin{equation}
\left( E_{\rm c} + E_{\rm p} \right)_t = - \nabla \cdot \left( \rho_0 p \textbf{u} \right) + D_{\rm diss} + \mathcal{P}_{\rm for},
\end{equation}

\noindent where we have introduced the kinetic energy \smc{per unit volume}, 

\begin{equation}
E_{\rm c} = \frac{1}{2} \rho_0 \textbf{u}^2,
\end{equation}

\noindent the potential energy \smc{per unit volume} associated with stratification,

\begin{equation}
E_{\rm p} = \frac{1}{2} \rho_0 \frac{b^2}{N^2}, 
\end{equation}

\noindent the power dissipated \smc{per unit volume} by the radiative cooling, 

\begin{equation}
D_{\rm diss} = \rho_0 \sigma_0 \frac{b^2}{N^2}
\end{equation}
 
\noindent and the power injected \smc{per unit volume} by the thermal tidal forcing, 

\begin{equation}
\mathcal{P}_{\rm for} = \rho_0 \frac{\kappa}{H N^2} b J.
\end{equation}

\noindent Averaged \jlc{over time}, \smc{these quantities} become

\begin{equation}
\begin{array}{ll}
   \displaystyle E_{\rm c} = \frac{1}{2} \rho_0 \langle \textbf{u}^2 \rangle_t, & 
   \displaystyle E_{\rm p} = \frac{1}{2} \frac{\rho_0}{N^2} \langle b^2 \rangle_t, \\[0.3cm]
   \displaystyle D_{\rm diss} = \frac{\rho_0 \sigma_0}{N^2} \langle b^2 \rangle_t , &
   \displaystyle \mathcal{P}_{\rm for} = \rho_0 \frac{\kappa}{H N^2} \langle b J \rangle_t, 
\end{array}
\end{equation}

\noindent where $ \langle \ldots \rangle_t $ \smc{is} the average in time. \rec{Finally}, we use the identity 

\begin{equation}
\langle \Re \left\{ p \right\}  \Re \left\{ q \right\} \rangle_t = \frac{1}{2} \Re \left\{ \pv{p} \, \pv{q}^* \right\},
\end{equation}

\noindent the notations $ \Re $ and $ * $ referring to the real part and conjugate of a complex number respectively. \smc{We} obtain 

\begin{equation}
\begin{array}{ll}
   \displaystyle E_{\rm c} = \frac{1}{4} \rho_0 \left| \pv{\textbf{u}} \right|^2, & 
   \displaystyle E_{\rm p} = \frac{1}{4} \frac{\rho_0}{N^2} \left| \pv{b} \right|^2, \\[0.3cm]
   \displaystyle D_{\rm diss} = \frac{1}{2} \frac{\rho_0 \sigma_0}{N^2} \left| \pv{b} \right|^2 , &
   \displaystyle \mathcal{P}_{\rm for} = \frac{1}{2} \frac{\rho_0 \kappa}{H N^2} \Re \left\{ \pv{b} \pv{J}^* \right\}. 
\end{array}
\label{energyz}
\end{equation}

\noindent \rec{We recall that}, in these expressions, $ \pv{b} $ and $ \pv{\textbf{u}} = \left( \pv{u} , \pv{v} , \pv{w} \right) $ designate the vertical profiles of the \jlc{complex} perturbed quantities that are given by Eqs.~(\ref{bz}-\ref{wz}). 

Similarly, the tidal torque exerted \rec{by the star} on the fluid layer with respect to the spin axis of the planet is defined by 

\begin{equation}
{\mathcal T}=\int_{\mathscr{V}} r sin \theta \Re \left\{ \delta  \rho \right\}  F_{T,\varphi} d \mathscr{V}
\end{equation}

\noindent \rec{where $ F_{\rm T ; \varphi} $ is the latitudinal component of the tidal force, $ \varphi $ the longitudinal coordinate and $ \mathscr{V} $ the volume of the fluid shell. This expression can be written as a function of the tidal potential generating the tidal force $ U $, which is such that $ \textbf{F} = \nabla U $, and becomes \cite[e.g.][]{Zahn1966a,ADLM2016a}}

\begin{equation}
\mathcal{T} = \Re \left\{ \frac{1}{2} \int_{\mathscr{V}} U_\varphi \delta \rho^* d \mathscr{V} \right\}.
\end{equation}

\noindent  \rec{One can note that $ \delta \rho $ has two components: the adiabatic component $ \Re \left\{ \delta \rho \right\} $, which is in phase with the tidal potential and does not contribute to the tidal torque, and the dissipative component delayed with respect to the forcing, $ \Im \left\{ \delta \rho \right\} $, the notation $ \Im $ referring to the imaginary part of a complex number.} Hence, the \smc{equivalent local} tidal torque exerted on the fluid box scales as

\begin{equation}
\mathcal{T} \propto - \Im \left\{ \int_0^{z_{\rm b}}  \rho_0 \pv{b} dz' \right\}.
\label{torque_local}
\end{equation}

\noindent

\section{\smc{Simplified atmospheric models}}
\label{sec:analytic_solutions}

The above equations are formulated in the general case, for any background distribution. \smc{In order to explore analytically the domain of parameters, we compute} in this section analytic solutions of the thermal tide in two simplified cases: in a homogeneous fluid with uniform background distributions and in an isothermal stably-stratified gas. In both cases, we consider a uniform profile of thermal forcing (\rec{$\pv{J}$ is now a constant}) for the sake of simplicity, \rec{as done before in the global model \citep{ADLM2016a}}. \rec{Such a profile physically corresponds to an optically thin atmosphere homogeneous in composition, where the stellar incoming flux is absorbed over the whole depth of the fluid layer. Is can also be seen in the general case as a zero-order approximation of the effective energy input per unit mass generating the thermal tide. Assuming $ \pv{J} $ to be constant allows us to reduce the right-hand member of the vertical structure equation (Eq.~\ref{vert_struct}) to a simple exponential function of the altitude.}

\subsection{In a homogeneous fluid}

The simplest case is \smc{that} of uniform background distributions, which corresponds to the case studied by \cite{GS2005}. We assume that $ \rho_0 $, $ H $ and $ \sigma_0 $ do not vary with the altitude. As a consequence, the vertical structure equation reduces to

\begin{equation}
\frac{d^2 \Psi}{dz^2} + k_\perp^2 \left[ \frac{\eta N^2 - \sigma^2}{\sigma^2 - f^2} + \left( \frac{\sigma f_{\rm s} }{\sigma^2 - f^2} \right)^2 \right] \Psi = \frac{k_\perp^2 \kappa \eta \pv{J}}{H \left( \sigma^2 - f^2 \right)} e^{- i \delta_{\rm c} z},
\label{vertical_hom}
\end{equation}

\noindent with 

\begin{equation}
\delta_{\rm c} = \frac{k_\perp f f_{\rm s}}{\sigma^2 - f^2}.
\end{equation}

Solving Eq.~(\ref{vertical_hom}) \rec{requires us} to choose two boundary conditions. As the guideline of this work is to characterize the atmospheric tidal response of a terrestrial planet, we use an \smc{\rec{impenetrable}} rigid-wall condition at $ z = 0 $, i.e. $\Psi = 0$. For the upper boundary, \smc{following \cite{SZ1990},} we assume that \smc{there is no material escape at the top of the atmosphere.} \smc{Thus,} the obtained solution shall not diverge at $ z = + \infty $. This amounts to eliminating the diverging term of the solution, 

\begin{equation}
\Psi = \mathscr{A} e^{i k_z z} + \mathscr{B} e^{-i k_z z} + \Psi_{\rm S} e^{- i \delta_{\rm c} z} ,
\end{equation}

\noindent where $ \Psi_{\rm S} $ is the amplitude of the particular solution, and $ \mathscr{A} $ and $ \mathscr{B} $ integration constants. With the convention $ \Im \left\{ k_z \right\} > 0 $, it follows that

\begin{equation}
\Psi = \mathscr{A} e^{i k_z z} + \Psi_{\rm S} e^{- i \delta_{\rm c} z},
\end{equation}

\noindent at the upper boundary. Thus, \rec{denoting by} $ \nu $ the dimensionless \jlc{complex} factor expressed as

\begin{equation}
\nu \left( \sigma \right) = \frac{\eta N^2}{\eta N^2 - \sigma^2 + f_{\rm s}^2},
\label{nu_uni}
\end{equation}

\noindent we end up with the solution

\begin{equation}
\Psi \left( z \right) = \Psi_{\rm S} \left( e^{- i \delta_{\rm c} z } - e^{i k_z z} \right),
\label{PsianaH}
\end{equation}

\noindent where the constant $ \Psi_{\rm S} $ \rec{is written}

\begin{equation}
\Psi_{\rm S} \left( \sigma \right) = \frac{\kappa \pv{J}}{H N^2} \nu .
\end{equation}

\noindent \rec{Note that the case $ \nu = 0 $ corresponds to a neutral stratification (see Eq.~(\ref{nu_uni}) with $ N^2 = 0 $). In this case, $ \Psi_{\rm S} = \kappa \pv{J} / \left[ H \left( f_{\rm s}^2 - \sigma^2 \right) \right] $.} We then substitute Eq.~(\ref{PsianaH}) in Eqs.~(\ref{bz}-\ref{wz}) and introduce the parameters

\begin{equation}
\begin{array}{lcl}
\displaystyle \mathcal{D} \left( z \right) = 1 - e^{i \left( \delta_{\rm c} + k_z \right) z}  & \mbox{and} &
 \displaystyle \mathcal{E} \left( z \right) = - i \left( \delta_{\rm c} + k_z \right) e^{i \left( \delta_{\rm c} + k_z \right)  z},
\end{array}
\label{paraDE}
\end{equation}

\noindent to get the vertical profiles of the tidal fluctuation of buoyancy

\begin{equation}
\pv{b} \left( z \right) = - \frac{i}{\sigma - i \sigma_0} \frac{\kappa \pv{J}}{H} \left( 1 - \nu \mathcal{D}  \right),
\label{bzh}
\end{equation}

\noindent pressure

\begin{equation}
\pv{p} \left( z \right) = i \frac{f^2 - \sigma^2}{\sigma k_\perp^2} \Psi_{\rm S} \left[ \mathcal{E}  + k_\perp \tilde{f} \frac{\sigma \cos \alpha + i f \sin \alpha}{f^2 - \sigma^2} \mathcal{D}  \right],
\end{equation}

\noindent and velocity field

\begin{align}
& \pv{u} \left (z \right) = i \frac{\cos \alpha}{k_\perp} \Psi_{\rm S} \left[ \mathcal{E}  + i k_\perp \frac{\tilde{f} \sin^2 \alpha}{i \sigma \cos \alpha + f \sin \alpha} \mathcal{D} \right], \\
& \pv{v} \left( z \right) = - \frac{f \cos \alpha - i \sigma \sin \alpha}{\sigma k_\perp} \Psi_{\rm S} \left[ \mathcal{E}  + i k_\perp \frac{\tilde{f} \sin \alpha \cos \alpha}{f \cos \alpha - i \sigma \sin \alpha} \mathcal{D}  \right], \\
& \pv{w} \left( z \right) = \Psi_{\rm S} \mathcal{D} .
\end{align}

\noindent Finally, denoting $ z_{\rm b} $ the altitude of the upper boundary and substituting Eq.~(\ref{bzh}) into Eqs.~(\ref{energyz}) \smc{and (\ref{torque_local})}, we obtain the total averaged dissipated energy per unit \smc{\rec{area}} of the $ \left( \sigma , k_\perp \right) $-mode in the fluid Cartesian box,

\begin{equation}
\hat{D}_{\rm diss} = \frac{1}{2} \frac{\rho_0 \sigma_0}{N^2 \left( \sigma^2 + \sigma_0^2 \right)} \left( \frac{\kappa \pv{J} }{H} \right)^2 \int_{0}^{z_{\rm b}} \left| 1 - \nu \mathcal{D} \left( z' \right) \right|^2 dz',
\label{Ddiss_uni}
\end{equation}

\noindent and the \rec{dimensionless} tidal torque 

\begin{equation}
\mathcal{T} \ \propto \ 2 \sigma_0 \Im \left\{ i \frac{1 - \nu}{\sigma - i \sigma_0} \right\},
\label{torque_uni}
\end{equation}

\noindent \rec{which is normalized so that $ \max \left| \mathcal{T} \right| = 1 $ when $ \nu = 0 $.}


\subsection{In an isothermal atmosphere}

The isothermal approximation is the zero-order approximation of the atmospheric structure of terrestrial planets such as the Earth. It is also the simplest structure allowing us to examine the repercussions of the variations of background distributions on the tidal response. Indeed, an isothermal atmosphere is characterized by exponentially decaying density and pressure distributions and a uniform pressure height-\smc{scale} depending on the temperature. Hence, $ \rho_0 $ is now given by

\begin{equation}
\begin{array}{lcl}
\displaystyle \rho_0 \left( z \right) = \rho_{\rm s} e^{- \tau z}, & \mbox{with} &
\displaystyle \tau = \frac{1}{H},
\end{array}
\end{equation}

\noindent the notations $ \rho_{\rm s} $ and $ \tau $ designating the density at the surface of the solid part of the planet and the \smc{vertical} decaying rate respectively. Since the density profile \smc{follows an exponential law}, $ d \ln \rho_0 / dz = - \tau $. Therefore, noticing that $ N^2 $ does not vary with the altitude either, we obtain for the vertical structure of tidal waves an equation of the same form as Eq.~(\ref{vertical_hom}), 

\begin{equation}
\frac{d^2 \Psi}{dz^2} + k_z^2 \Psi = \frac{k_\perp^2 \kappa \eta \pv{J}}{H \left( \sigma^2 - f^2 \right)} e^{- i \delta_{\rm c} z},
\end{equation}

\noindent with \smc{in this case}

\begin{equation}
\delta_{\rm c} = \frac{k_\perp f f_{\rm s}}{\sigma^2 - f^2} - i \frac{\tau}{2}
\end{equation}

\noindent and 

\begin{equation}
k_z^2 = k_\perp^2 \left[ \frac{\eta N^2 - \sigma^2}{\sigma^2 - f^2} + \left( \frac{\sigma f_{\rm s}}{\sigma^2 - f^2} \right)^2 \right] - \tau  k_\perp \frac{\sigma \tilde{f} \cos \alpha}{\sigma^2 - f^2} + \frac{\tau^2}{4}.
\label{kz2_iso}
\end{equation}

\noindent Hence, by applying the same boundary conditions as in the previous case, we obtain an analytic solution, \smc{which} is written similarly as Eq.~(\ref{PsianaH}). The parameter $ \nu $ is modified solely and \rec{is written}

\begin{equation}
\nu  = \frac{\eta N^2}{\eta N^2 - \sigma^2 + f_{\rm s}^2 + \gamma \tilde{f} \left( i f \sin \alpha - \sigma \cos \alpha \right) + \frac{1}{2} \gamma^2 \left( \sigma^2 - f^2 \right)  },
\label{nu_iso}
\end{equation}

\noindent where \rec{the parameter $ \gamma $ is defined as} 

\begin{equation}
\gamma = \frac{\tau}{k_\perp},
\end{equation}

\noindent \rec{and} compares the horizontal wavelength of the mode to the typical scale height of the background \smc{vertical} distributions. The vertical profiles of perturbed quantities derived from this solution are written for the buoyancy

\begin{equation}
\pv{b} \left( z \right) = - \frac{i}{\sigma - i \sigma_0} \frac{\kappa \pv{J}}{H} \left( 1 - \nu \mathcal{D} \right),
\end{equation}

\noindent for the pressure

\begin{equation}
\pv{p} \left( z \right) = i \frac{f^2 - \sigma^2}{\sigma k_\perp^2} \Psi_{\rm S} \left[ \mathcal{E}  + \left( k_\perp \tilde{f} \frac{\sigma \cos \alpha + i f \sin \alpha}{f^2 - \sigma^2} - \tau \right) \mathcal{D}  \right],
\label{pressure_iso}
\end{equation}

\noindent and for the velocity field

\begin{align}
 \pv{u} \left (z \right) = & \, i \frac{\cos \alpha}{k_\perp} \Psi_{\rm S} \left[ \mathcal{E}  + \left( i k_\perp \frac{\tilde{f} \sin^2 \alpha}{i \sigma \cos \alpha + f \sin \alpha} - \tau \right) \mathcal{D} \right], \\[0.3cm]
 \pv{v} \left( z \right) = & - \frac{f \cos \alpha - i \sigma \sin \alpha}{\sigma k_\perp} \Psi_{\rm S} \left[ \mathcal{E}  \right. \\
  & \left. + \left( \frac{ i k_\perp \tilde{f} \sin \alpha \cos \alpha}{f \cos \alpha - i \sigma \sin \alpha}- \tau \right) \mathcal{D}  \right], \nonumber \\[0.3cm]
 \pv{w} \left( z \right) = & \, \Psi_{\rm S} \mathcal{D} .
\end{align}

In these expressions, the parameters $ \mathcal{D} $ and $ \mathcal{E} $ are those defined by Eq.~(\ref{paraDE}). Basically, it shall be noted that we recover all of the results of the previous case by setting $ \tau = 0 $ since a uniform density distribution is just the asymptotic limit of an exponentially decaying one with $ H \rightarrow + \infty $. We finally compute the energy dissipated per unit \smc{surface} by \smc{the radiative cooling} in the fluid section

\begin{equation}
\hat{D}_{\rm diss} = \frac{1}{2} \frac{\rho_{\rm s} \sigma_0}{N^2 \left( \sigma^2 + \sigma_0^2 \right)} \left( \frac{\kappa \pv{J}}{H} \right)^2 \int_0^{z_{\rm b}} \left| 1 - \nu \mathcal{D} \left( z' \right) \right|^2 e^{- \tau z'} dz',
\label{Ddiss_iso}
\end{equation}

\noindent and of the \rec{normalized dimensionless} tidal torque,

\begin{equation}
\mathcal{T} = 2 \sigma_0 \Im \left\{ \frac{i}{\sigma - i \sigma_0} \left( 1 - \nu - \frac{\nu}{1 - i \frac{\delta_{\rm c} + k_z}{\tau}} \right) \right\},
\label{torque_iso}
\end{equation}


\noindent which \rec{is obtained in a similar way as Eq.~(\ref{torque_uni}) and} is such that $ \max \left| \mathcal{T} \right| = 1 $ when $ \nu = 0 $. As noted for other quantities, we easily verify that these two expressions \smc{simplify} into those obtained in the case of uniform background distributions, i.e. Eqs.~(\ref{Ddiss_uni}) and (\ref{torque_uni}) respectively, when $ \tau \rightarrow 0 $.

\section{Tidal regimes and their implications on the rotational dynamics of terrestrial planets}
\label{sec:tidal_regimes}

 In this section, we use the results derived in the framework of \smc{our} Cartesian \smc{model} to explore the domain of parameters and understand the behaviour of the atmospheric tidal response \smc{predicted} by the much more \smc{complex previous} global modelings \citep[][]{ADLM2016a,Leconte2015}. Particularly, we examine the tidal torque exerted on the atmosphere, which contributes to the evolution of the rotation rate of the planet with the torque resulting from the tidal elongation \smc{and the viscous friction} of the solid \smc{core}. In the vicinity of \smc{synchronization}, the atmospheric and solid \smc{tidal torques} can balance each other, the first one \smc{torquing} the planet \smc{\textit{away} \rec{from} synchronous rotation} and the second one \smc{\textit{towards} \rec{it}}. This \smc{can} explain the locking of the \rec{planet Venus} at the observed \smc{asynchonous} retrograde rotation rate \citep[][]{GS1969,ID1978,DI1980,CL01,CL2003}.
 
  If the tidal response of the atmosphere is reduced to its \emph{non-wavelike} part\footnote{\smc{The separation of the fluid tidal response into \emph{non-wavelike} and \emph{wavelike} part was introduced by \cite{Ogilvie2013} in the context of gravitational tides. The non-wavelike part designates the instantaneous global elongation of the atmosphere \smc{and the induced large-scale flow}, while the wave-like part encompasses the effects of internal waves generated by the forcing and resonances they can induce. In the present work, the term non-wavelike refers to the global elongation of the atmosphere due to the thermal forcing by analogy with the gravitational tidal potential. It is however different from the \emph{equilibrium tide}, which designates the tidal response of the fluid in the zero-frequency limit ($ \sigma \rightarrow 0 $).}} \smc{\citep[][]{Ogilvie2013}}, which is associated with a quadrupolar bulge, the atmospheric tidal torque can be approximated with a Maxwell rheology characterized by the effective radiative frequency of the atmosphere \citep[][]{ID1978,Leconte2015,ADLM2016a,ADLM2016b}, i.e. 

\begin{equation}
\mathcal{T} \ \propto \ \frac{\sigma}{\sigma^2 + \sigma_0^2}.
\label{torque_Maxwell}
\end{equation}

However, in \cite{ADLM2016a}, we pointed out the fact that stable stratification \smc{is} able to annihilate the tidal torque due to the thermal \smc{non-wavelike tide} in the vicinity of synchronization \smc{because it induces} a \emph{wavelike tide} composed of gravity waves. As a consequence, the planet would be led in this case towards synchronous rotation by the solid tide. \smc{To study this effect,} the global modeling used in this early study \smc{should be completed for two reasons}. First, it describes the tidal atmospheric response in a spherical geometry. \smc{This tends to dilute the physics into} mathematical \smc{aspects}. Second, it is based upon the traditional approximation, and can be affected by the bias induced by this hypothesis, \smc{which will be} studied in the next section. \smc{Typically, when $ 2 \Omega \gtrsim N $, the traditional approximation \rec{can lead us to} overestimate the amplitude of the tidal torque, as discussed by \cite{OL2004}. These authors show that the \emph{static approximation}, \smc{in which} the effect of rotation is ignored, is better in this case. Following \smc{them}, we used this \smc{last} approximation instead of the traditional approximation to establish the Maxwell law in the global model \citep[][]{ADLM2016a}}. 

Hence, the local Cartesian modeling developed in the present study will allow us \smc{to explore the full domain of parameters in rotation and stratification with taking into account the complete Coriolis acceleration, filtering out complexities due to the spherical geometry but keeping the key \smc{physical ingredients}.}

\begin{figure}
   \centering
   \includegraphics[width=0.48\textwidth]{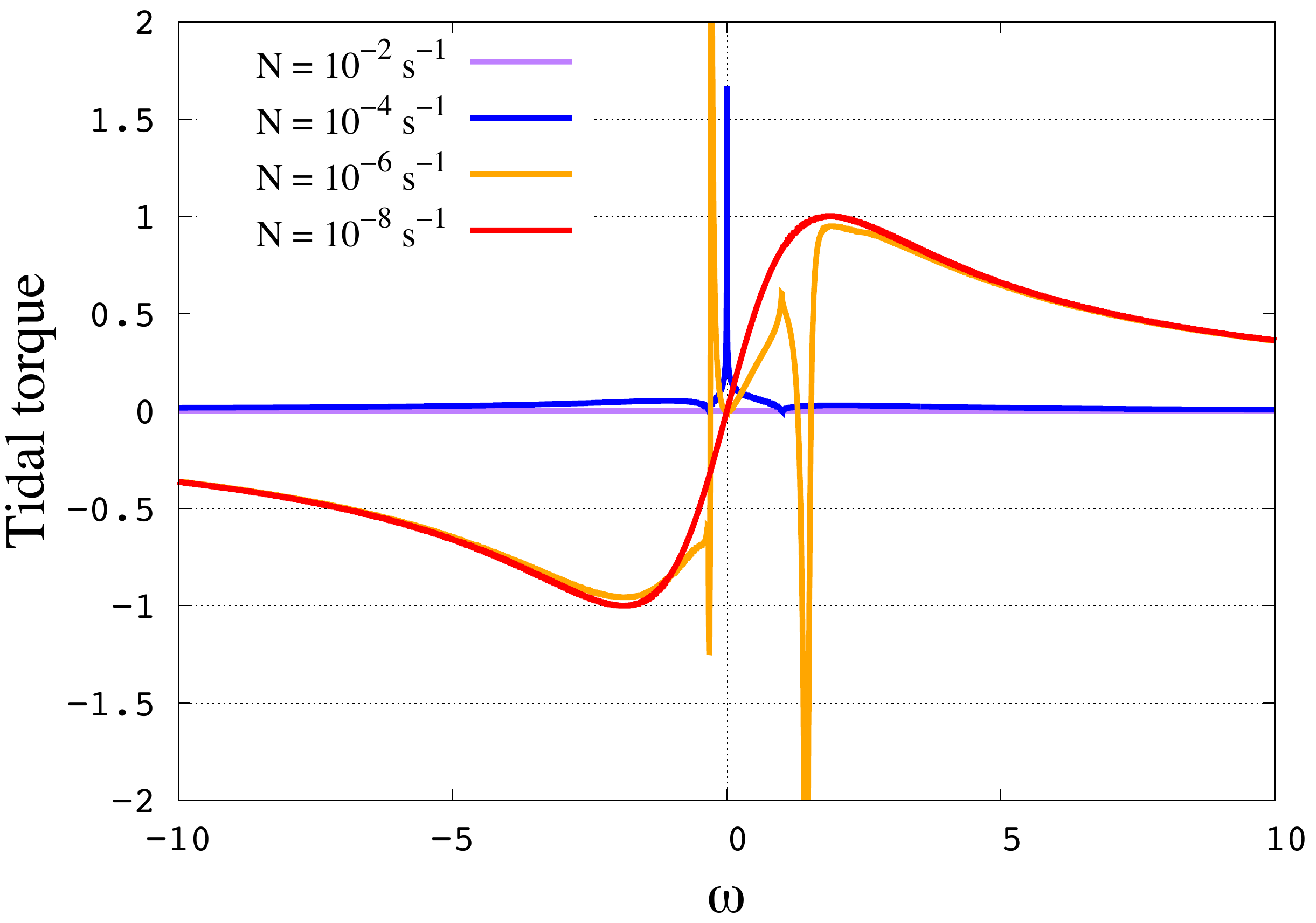}
   \caption{\rec{Normalized tidal torque} as a function of the frequency of the perturber $ \omega = \left( \Omega - n_{\rm orb} \right)/ n_{\rm orb} $ for various values of the Brunt-Väisälä frequency from weak to strong stable stratification, i.e. \smc{$ \log \left( N \right) = \left\{ -8,-6,-4,-2 \right\} $}. The tidal torque is computed using the normalized function given by Eq.~(\ref{torque_iso}) and we consider the case of a Venus-like planet with the following set of values: $ n_{\rm orb} = 1.991 \times 10^{-7} \ {\rm s^{-1}} $, $ \Gamma_1 = 1.4 $ (perfect gas), $ H = 15.9 $ km, $ \theta = \pi/3 $, $ \alpha = 0 $ (westward propagating wave), $ \sigma_0 = 7.5 \times 10^{-7} \ {\rm s^{-1}} $ \citep[][]{Leconte2015}, and $ k_\perp = 2 \pi / \lambda $ with $ \lambda = 1000 $~km.}
\label{fig:couple_N2_1D}%
\end{figure}

Considering the expressions of $ \nu $ in the two cases treated in the previous section, Eqs.~(\ref{nu_uni}) and (\ref{nu_iso}), we recover the asymptotic regimes observed with global modelings. On the one hand, in the convective atmosphere limit (i.e. $ N^2 \rightarrow 0 $), $ \nu \rightarrow 0 $. The expressions of the tidal torque given by Eqs.~(\ref{torque_uni}) and (\ref{torque_iso}) both reduce to Eq.~(\ref{torque_Maxwell}), i.e. \smc{a} Maxwell \smc{law}. On the other hand, in the \smc{stably}-stratified atmosphere limit ($ N^2 \rightarrow + \infty $), $ \nu \rightarrow 1 $. Thus, the torque obtained in the \rec{case of a uniform} background readily tends to zero. The case of the isothermal atmosphere is a little bit more \smc{complex} because we have to take into account the dependence of the vertical wavelengths of tidal gravito-inertial waves on the tidal frequency. Eq.~(\ref{kz2_iso}) shows that \jlc{$ k_z \left( \sigma \right) $} diverges in the vicinity of spin-orbit synchronization ($ \sigma = 0 $), namely $ k_z \ \propto \ \left| \sigma^2 - f^2 \right|^{-1/2} $ if $ \left| \sigma \right| \gg \sigma_0 $, and $ k_z \ \propto \ \left| \sigma / \left( \sigma^2 - f^2 \right) \right|^{1/2} $ otherwise. As a consequence, $ \left| k_z/\tau \right| \gg 1 $ and the \smc{third} term associated with the propagation of gravito-inertial waves in Eq.~(\ref{torque_iso}) becomes negligible. We \smc{identify} here how the tidal torque is \emph{flattened} by the stable stratification. \smc{This behaviour corresponds to the equilibrium thermal tide of a stably stratified fluid region studied by \cite{AS2010} in the case of fluid extrasolar planets. As demonstrated analytically by \cite{AS2010}, the $ N^2 \xi_r / g $ term of the heat transport equation tends to equalize the heat source term ($ J $) in the equation of energy (Eq.~(\ref{prim_buoyancy})) while $ \sigma \rightarrow 0 $. This means that \rec{the local density decreases} generated by the thermal forcing are exactly cancelled by the vertical displacement of denser fluid brought up from below.}

\begin{figure*}
   \centering
   \includegraphics[width=0.42\textwidth,trim = 1.0cm 4.0cm 2.5cm 2.0cm, clip]{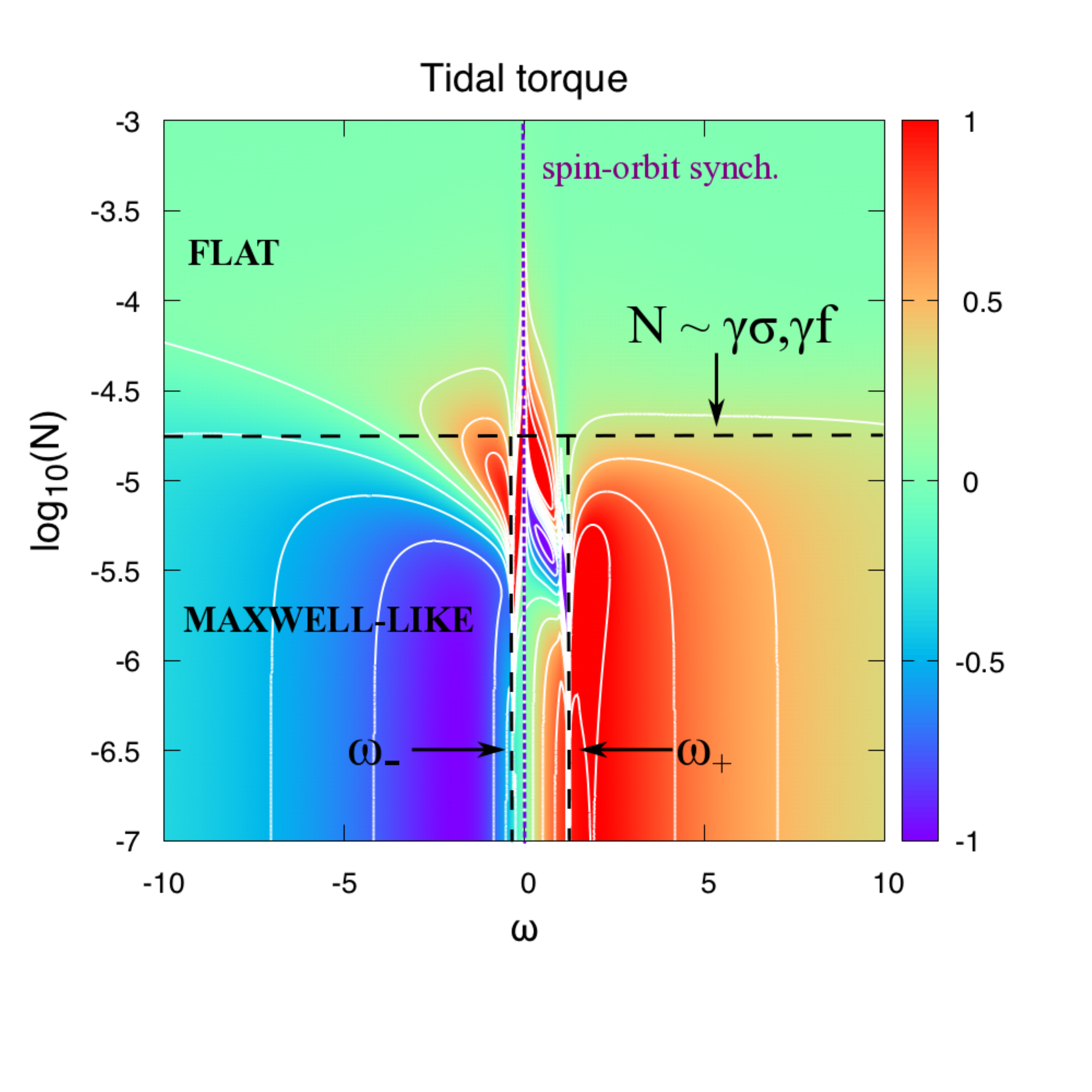} \hspace{1cm}
   \includegraphics[width=0.42\textwidth,trim = 1.0cm 4.0cm 2.5cm 2.0cm, clip]{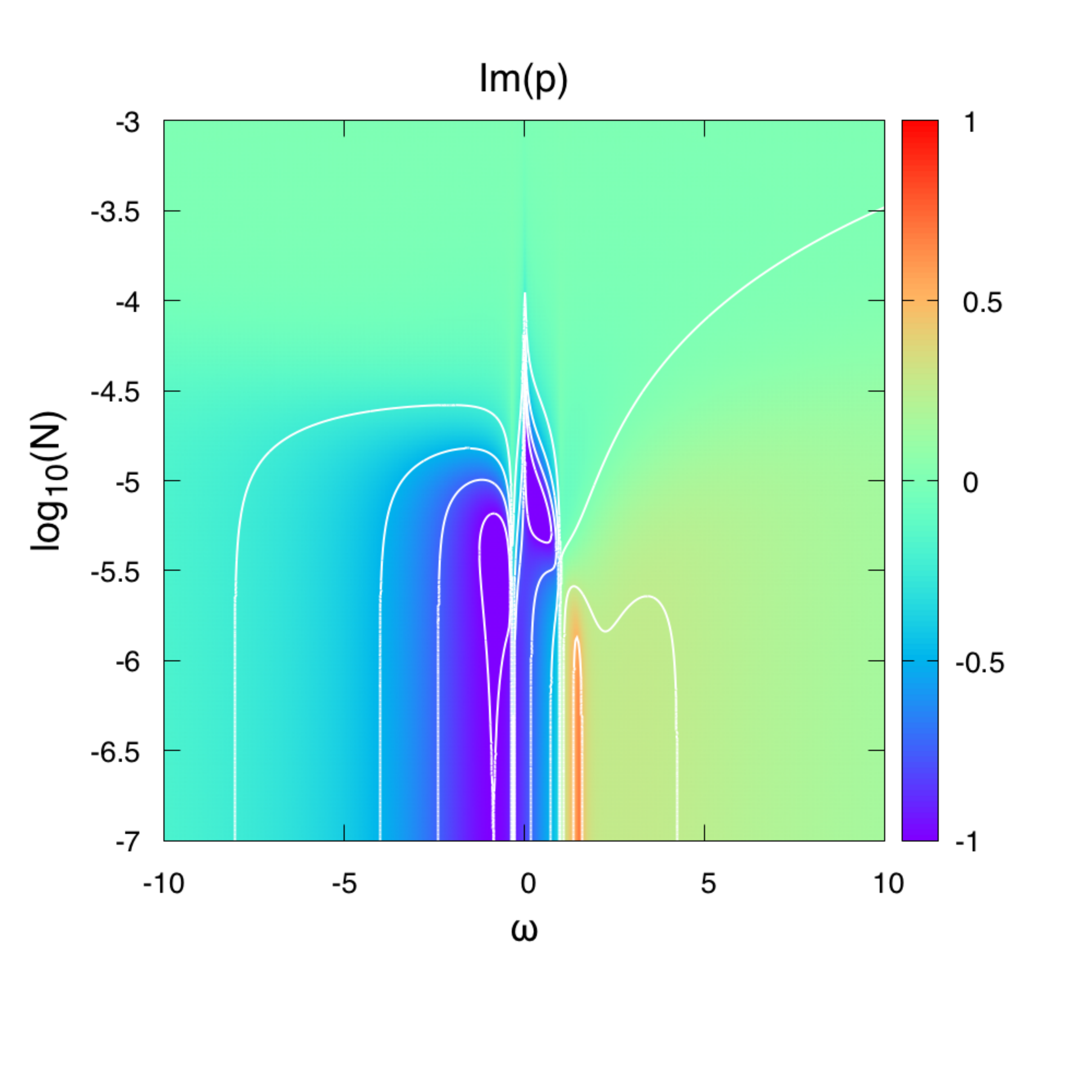}
   \caption{\jlc{Normalized tidal torque (left panel) and imaginary part of the surface pressure oscillation (right panel) as functions of the normalized tidal frequency $ \omega = \left( \Omega - n_{\rm orb} \right) / n_{\rm orb} $ (horizontal axis) and Brunt-Väisälä frequency in logarithmic scale (vertical axis). The tidal torque and pressure variations are computed using the functions given by Eqs.~(\ref{torque_iso}) and (\ref{pressure_iso}) respectively. \smc{The torque is normalized by its maximum value in the asymptotic regime of neutral stratification.} The imaginary part of pressure oscillations is normalized by its maxima in absolute value.} We consider the case of a Venus-like planet with the following set of values: $ n_{\rm orb} = 1.991 \times 10^{-7} \ {\rm s^{-1}} $, $ \Gamma_1 = 1.4 $ (perfect gas), $ H = 15.9 $ km, $ \theta = \pi/3 $, $ \alpha = 0 $ (westward propagating wave), $ \sigma_0 = 7.5 \times 10^{-7} \ {\rm s^{-1}} $ \citep[][]{Leconte2015}, and $ k_\perp = 2 \pi / \lambda $ with $ \lambda = 1000 $~km. \smc{The frequencies $ \omega_- $ and $ \omega_+ $ are the resonant frequencies identified in Section~\ref{sec:tidal_regimes} (see Eq.~(\ref{omegapm})).} }
\label{fig:couple_map}%
\end{figure*}

This behaviour is illustrated by Figs.~\ref{fig:couple_N2_1D} and \ref{fig:couple_map} where the case of the stellar semidiurnal thermal tide is examined. The planet is supposed to orbit its host star circularly at the orbital frequency $ n_{\rm orb} $, and its equatorial plane is coplanar with the orbital plane, so that the perturbation reduces to the quadrupolar tidal forcing of frequency $ \sigma = 2 \left( \Omega - n_{\rm orb} \right) $. We set for the studied mode $ k_\perp = 2 \pi / \lambda $, with the wavelength $ \lambda = 1000 $ km, so that \smc{$ \gamma = \tau / k_\perp \gg 1 $} (asymptotic regime of large wavelengths). The tidal torque is plotted as a function of \rec{the normalized apparent orbital} frequency of the star in the reference frame in co-rotation with the planet, namely \rec{the normalized tidal frequency} $ \omega = \left( \Omega - n_{\rm orb} \right) / n_{\rm orb} $, for different values of the Brunt-Väisälä frequency from the convective \smc{isentropic} limit ($ N = 10^{-8} \ {\rm s^{-1}} $) to strong stable stratification ($ N = 10^{-2} \ {\rm s^{-1}} $). \smc{These boundaries are chosen in such way that $ N \ll \left\{ 2 \Omega, \sigma , \sigma_0 \right\} $ in the first asymptotic regime and $ N \gg \left\{ 2 \Omega , \sigma , \sigma_0 \right\} $ in the other one.} We recover in the first case (red curve) the Maxwell-like tidal response predicted by early studies \cite[][]{ID1978,CL01,ADLM2016a}, and in the second one (violet curve) the weak torque obtained with the ab initio modeling of \cite{ADLM2016a} \smc{that takes tidal gravity waves into account}. 

Several resonances can be observed in the transition regime. They result directly from the \smc{equality} of characteristic frequencies of the system (\rec{$N^2 $}, $ \sigma $, $ 2 \Omega $), which occurs in the denominator of the parameter $ \nu $. \rec{They are related to the local nature of the model since their positions depend on the colatitude, and consequently do not exist in the global tidal response, where the tidal equation is integrated over the sphere \citep[][]{ADLM2016a}. However, we have to characterize them here to clarify the tidal torque frequency spectra observed in Fig.~\ref{fig:couple_N2_1D}.}

In the regime of large wavelengths ($ \gamma \gg 1 $), the denominator of Eq.~(\ref{nu_iso}) can be approximated by 

\begin{equation}
d = \eta N^2 + \frac{1}{2} \gamma^2 \left( \sigma^2 - f^2 \right). 
\label{den1}
\end{equation}

\noindent Let us assume $ \sigma_0 \lesssim \sigma $, \smc{which corresponds to the} quasi-adiabatic regime \smc{\citep[][]{Press1981,ADMLP2015}}, and substitute $ \sigma = 2 \left( \Omega - n_{\rm orb} \right) $ in Eq.~(\ref{den1}). Resonances thus correspond to the zeros of the polynomial 

\begin{equation}
P \left( \Omega \right) = \sin^2 \theta \, \Omega^2 - 2 n_{\rm orb} \Omega + n_{\rm orb}^2 + \frac{1}{2} \gamma^{-2} N^2.
\end{equation}

\noindent We \smc{obtain} the roots 

\begin{equation}
\Omega_{\pm} = \frac{1}{\sin^2 \theta} \left[ n_{\rm orb} \pm \sqrt{n_{\rm orb}^2 \cos^2 \theta - \frac{1}{2} \gamma^{-2} N^2 \sin^2 \theta } \right],
\end{equation}

\noindent which, expressed in the normalized tidal frequency $ \omega = \left( \Omega - n_{\rm orb} \right) / n_{\rm orb} $ \smc{used in} Fig.~\ref{fig:couple_N2_1D}, become

\begin{equation}
\omega_{\pm} = \frac{\cos \theta}{\sin^2 \theta} \left[ \cos \theta \pm \sqrt{ 1 - \frac{1}{2} \gamma^{-2} \left( \frac{N}{n_{\rm orb}} \right)^2 \tan^2 \theta }  \right]. 
\label{omegapm}
\end{equation}

\noindent In the limit $ N^2 \rightarrow 0 $, they tend to $ \omega_{-} = \tan^{-2} \theta - 1 $ and $ \omega_{+} = \sin^{-2} \theta $. As a consequence, $ \omega_- = -1 $ and $ \omega_+ = + \infty  $ at the poles. 

These features are represented on the right panel of Fig.~\ref{fig:couple_map} where the tidal torque is plotted as a function of the tidal frequency ($ \omega $) and Brunt-Väisälä frequency in logarithmic scale. Black vertical dashed lines indicate the positions of resonances, while the horizontal dashed line designates the transition between the \emph{flat} and \emph{Maxwell-like} regimes \smc{for the tidal torque} identified above, which corresponds to $ N \sim \left\{ \gamma \sigma , \gamma f \right\} $. \smc{For} the evolution of planetary systems, this means that rocky planets with an \rec{atmospheric} \jlc{Brunt-Väisälä frequency} below this critical \jlc{value} are likely to tend towards non-synchronized rotation states of equilibrium, like Venus, while those beyond it will be led towards spin-orbit synchronization, in good agreement with \smc{predictions obtained in} \cite{ADLM2016a} \jlc{(Section~6.3)}. 

\jlc{\rec{In} the right panel of Fig.~\ref{fig:couple_map}, the imaginary part of the surface pressure oscillation is plotted as a function of $ \omega $ and $ N $. This plot shows that we retrieve the asymptotic behaviours identified for the tidal torque in surface pressure oscillations. In a weakly stratified atmosphere, a net tidal bulge appears, leading to surface pressure oscillations of high amplitude in agreement with the lag of the bulge. In the strongly stratified regime, pressure oscillations vanish as there is no net tidal bulge anymore. However, we note that $ \Im \left\{ \pv{p} \right\} $ has not the same functional form as the tidal torque. Besides, one should bear in mind the fact that it partly depends on the chosen boundary conditions, which prevents us to proceed to a more quantitative analysis. }

To illustrate \jlc{the impact of the atmospheric structure on the long-term rotational evolution,} we study the evolution of the rotation rate of \smc{an idealized} Venus-like rocky planet submitted to both atmospheric and solid semidiurnal tides \smc{due to a host star}. \smc{As the goal of these calculations is to isolate the different possible evolutions, we choose simple values of parameters.} We use for the solid torque the simplified model expressed as 

\begin{equation}
\mathcal{T}_{\rm S} = - \mathcal{T}_{\rm S ; 0} \tanh \left( \frac{\sigma}{\sigma_{\rm S}} \right),
\label{torque_solid}
\end{equation}

\noindent where $ \mathcal{T}_{\rm S ; 0} = 5.0 \times 10^{16} $ N.m designates the amplitude of the torque and $ \sigma_{\rm S} $ the effective relaxation frequency of the material composing the \smc{rocky core}. This model can be interpreted as a zero-order approximation of the Andrade model \citep[][]{Andrade1910,Efroimsky2012,Leconte2015}, which describes the forced visco-elastic response of metals and silicates. The relaxation frequency of a rocky planet is usually far smaller than the orbital frequency \citep[$ \sigma_{\rm S} \sim 10^{-10} \ {\rm s^{-1}} $, e.g.][]{Efroimsky2012}. Here, we set $ \sigma_{\rm S} = 10^{-2} \, n_{\rm orb}$. In the Andrade model, the decaying rate of the tidal torque is far smaller than the one of the atmosphere for $ \sigma \gg \sigma_{\rm S} $. So, it is well approximated by a constant in the vicinity of synchronous rotation. \smc{Concerning} the atmospheric torque, we apply Eq.~(\ref{torque_iso}) with the effective radiative frequency $ \sigma_0 = 3 \ n_{\rm orb} $ and the amplitude $ \mathcal{T}_{\rm A ; 0} = 1.52 \ \mathcal{T}_{\rm S ; 0} $, which is \smc{arbitrarily} chosen in such way that the final rotation rate corresponds to $ \omega = 4 $ in case of asynchronous state of equilibrium. The planet is assumed to have the same moment of inertia as Venus, i.e. $ \mathcal{I}_{\rm p} = 5.88 \times 10^{37} \ {\rm kg.m^2} $, and to orbit its host star at the orbital period $ P_{\rm orb} = 100 $ d (\rec{let us recall} that $ n_{\rm orb} = 2 \pi / P_{\rm orb} $).  

\smc{The} equation describing the evolution of the planet's rotation rate \rec{is written}

\begin{equation}
\mathcal{I}_{\rm p} \frac{d \Omega}{dt} = \mathcal{T}_{\rm S} \left( \Omega \right) + \mathcal{T}_{\rm A} \left( \Omega \right).
\end{equation}

\begin{figure}
   \centering
   \includegraphics[width=0.49\textwidth]{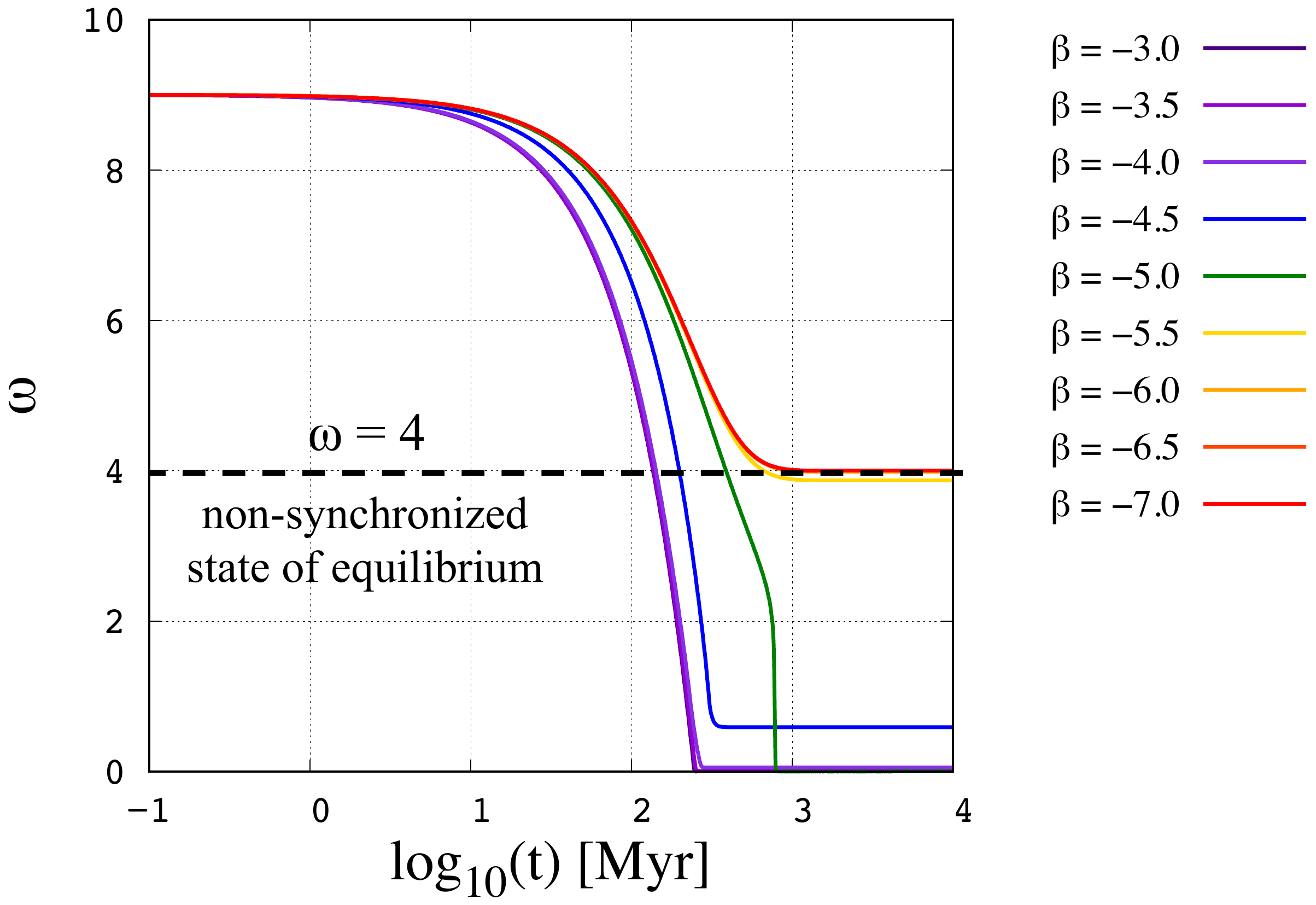}
   \caption{Evolution of the rotation rate of a Venus-like planet for various Brunt-Väisälä frequencies. The normalized frequency $ \omega = \left( \Omega - n_{\rm orb} \right) / n_{\rm orb} $ is plotted as a function of time (Myr) in logarithmic scale. The Brunt-Väisälä frequency \jlc{$ N = 10^\beta $} is \smc{increased} from $ \beta = - 7 $ (weak stable stratification) to $ \beta = -3 $ (strong stable stratification) with \smc{a} step $ \Delta \beta = 0.5 $.  }
\label{fig:evolution}%
\end{figure}

\noindent It is integrated over 10 billion years using the ODEX code implemented in the algebraic manipulator TRIP \citep[see][]{Hairer1993,GL13}, with the initial condition $ \Omega \left( 0 \right) = 10 \ n_{\rm orb} $. Simulations are achieved for a large range of Brunt-Väisälä frequencies \jlc{$ N = 10^\beta $} with $ \beta = -7 $ to $ \beta = -3 $. Results are plotted \rec{in} Fig.~\ref{fig:evolution} as a function of time.

We can observe \rec{in this figure} that \smc{rotational evolutions} divide into two distinct families. The first family \smc{($\beta \geq -4$)}, which encompasses strongly-stratified cases, is driven by the solid tidal torque solely. In this family, the rotation rate invariably converges towards synchronous rotation ($ \omega = 0 $). Planets of the second family \smc{($\beta \leq -5.5$)} are driven by both  solid and atmospheric tidal torques. The evolution rate of their spin varies depending on the strength of the atmospheric tidal torque, which tends to push the planet \emph{away} from synchronous rotation. They finally reach a stable non-synchronized state of equilibrium corresponding to the frequency where the solid and atmospheric torques exactly balance each other. We can notice two \smc{evolutions} behaving in a different way than those detailed above, namely $ \beta = - 4.5 $ and $ \beta = - 5$. These cases are associated with the transition regime. In the first one ($ \beta = - 4.5 $) a non-synchronized state of equilibrium closer to synchronization is reached, while the rapid evolution of $ \Omega $ in the second one ($ \beta = -5  $) around $ t = 600 $ Myr results from one of the resonances studied above (see Eqs~(\ref{den1}) to (\ref{omegapm})). \smc{We note that the rotation tends towards synchronization in this case.}

\smc{ In the general case, it seems difficult to predict the final rotation rate of the planet using the formula given by Eq.~(\ref{torque_iso}). However, it can be done in the asymptotic regime described by the Maxwell model \citep[see e.g.][in the case where the tidal torque of the rocky core is modeled by a Maxwell law]{ADLM2016b}. With the law chosen for \smc{the tidal torque applied to} the rocky core in the present work (Eq.~\ref{torque_solid}), the frequencies $ \omega_\pm $ of possible states of equilibrium are expressed as}

\begin{equation}
\omega_\pm = \frac{1}{2} \frac{\sigma_0}{n_{\rm orb}} \frac{\mathcal{T}_{\rm A ; 0}}{\mathcal{T}_{\rm S ; 0}} \left[ 1 \pm \sqrt{1 - \left( \frac{\mathcal{T}_{\rm S ; 0}}{\mathcal{T}_{\rm A ; 0}} \right)^2} \right],
\end{equation}

\noindent \smc{although \rec{only} $ \omega_+ $ corresponds to a stable state leading to an asynchronous final rotation rate. }

\begin{figure*}
   \centering
   \hspace{1.0cm}
     \includegraphics[height=0.03\textwidth]{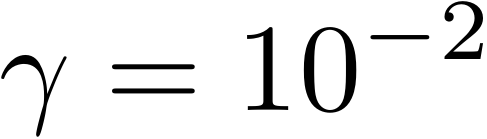} \hspace{4cm}
   \includegraphics[height=0.03\textwidth]{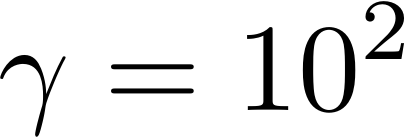} \\
   \raisebox{1.5cm}{\includegraphics[width=0.03\textwidth]{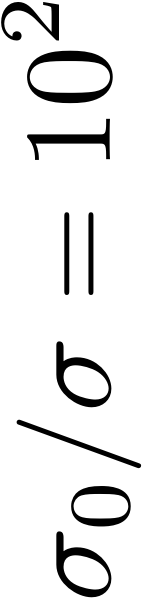}}
   \hspace{0.1cm}
   \raisebox{2.0cm}{\includegraphics[width=0.03\textwidth]{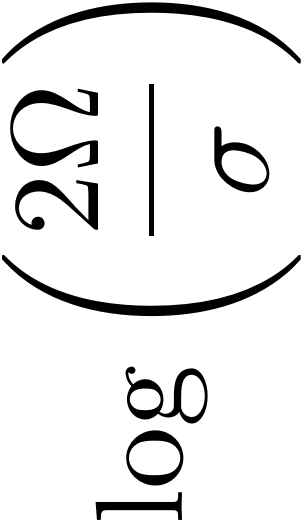}}
   \includegraphics[width=0.31\textwidth,trim = 2.5cm 5.9cm 2.5cm 3.4cm, clip]{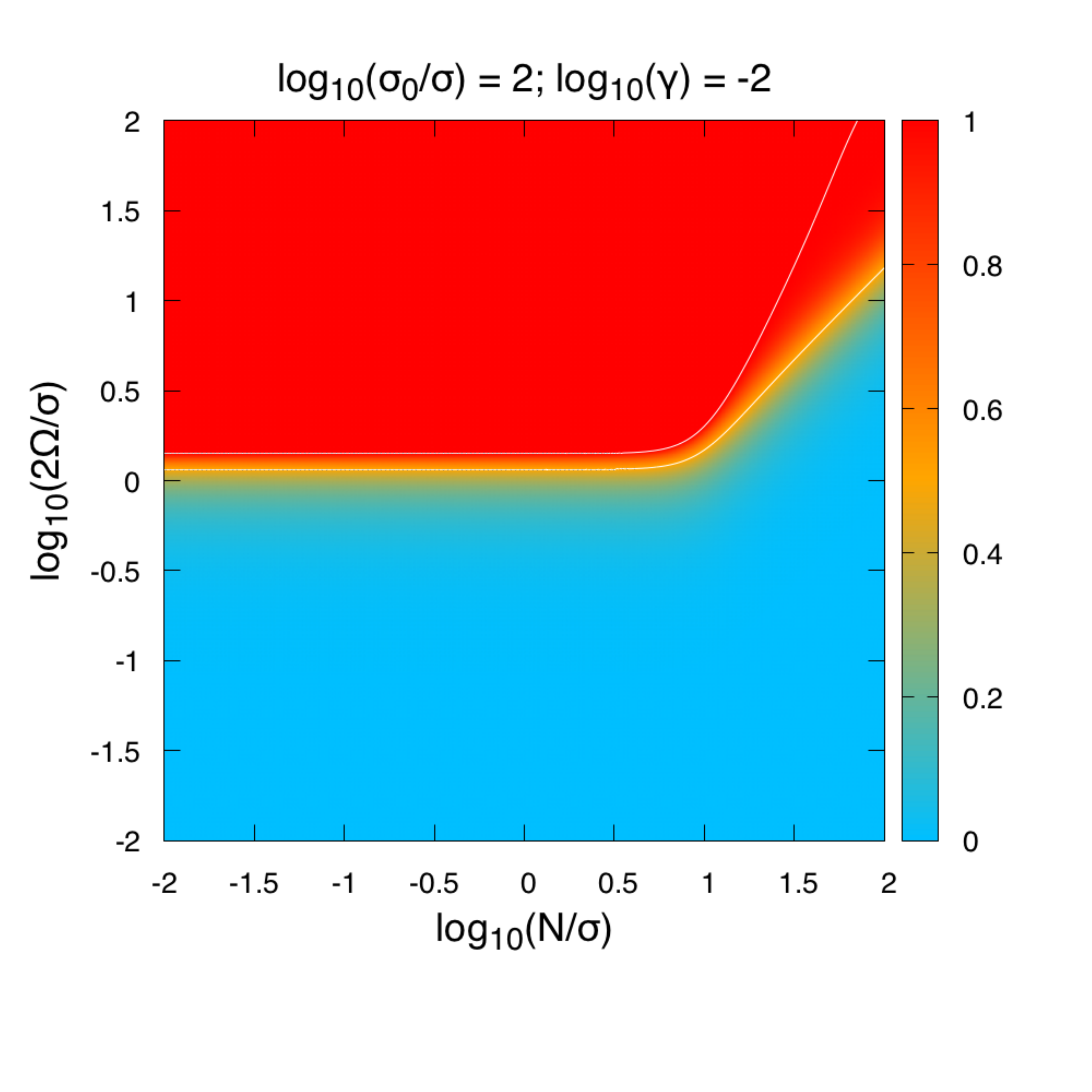}
   \includegraphics[width=0.31\textwidth,trim = 2.5cm 5.9cm 2.5cm 3.4cm, clip]{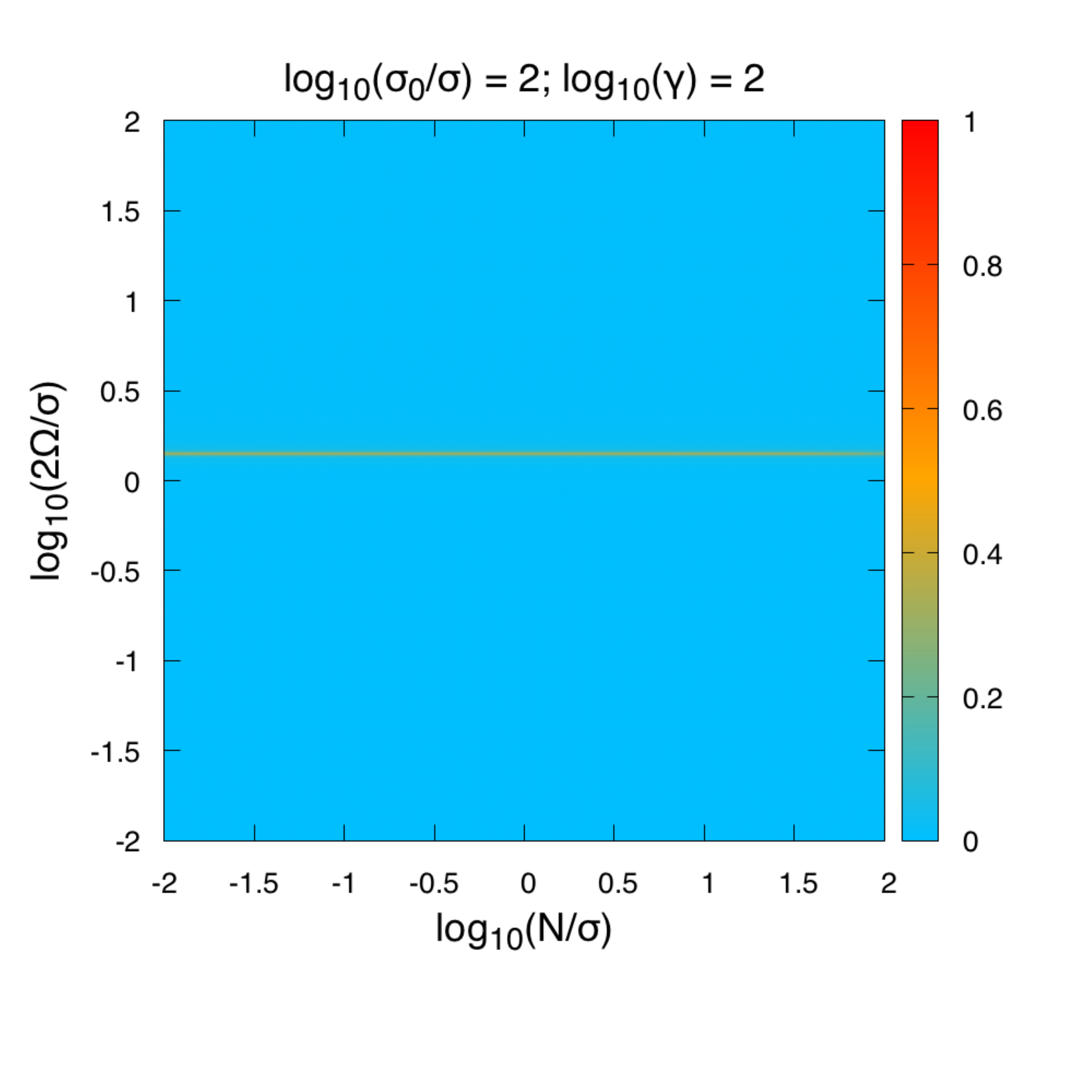} \\
   \raisebox{1.5cm}{\includegraphics[width=0.03\textwidth]{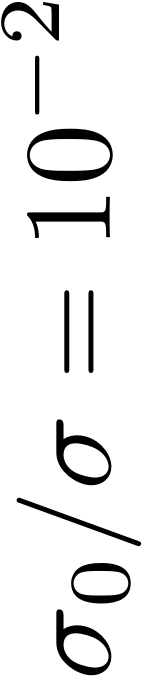}}
   \hspace{0.1cm}
   \raisebox{2.0cm}{\includegraphics[width=0.03\textwidth]{auclair-desrotour_fig5d.pdf}}
   \includegraphics[width=0.31\textwidth,trim = 2.5cm 5.9cm 2.5cm 3.4cm, clip]{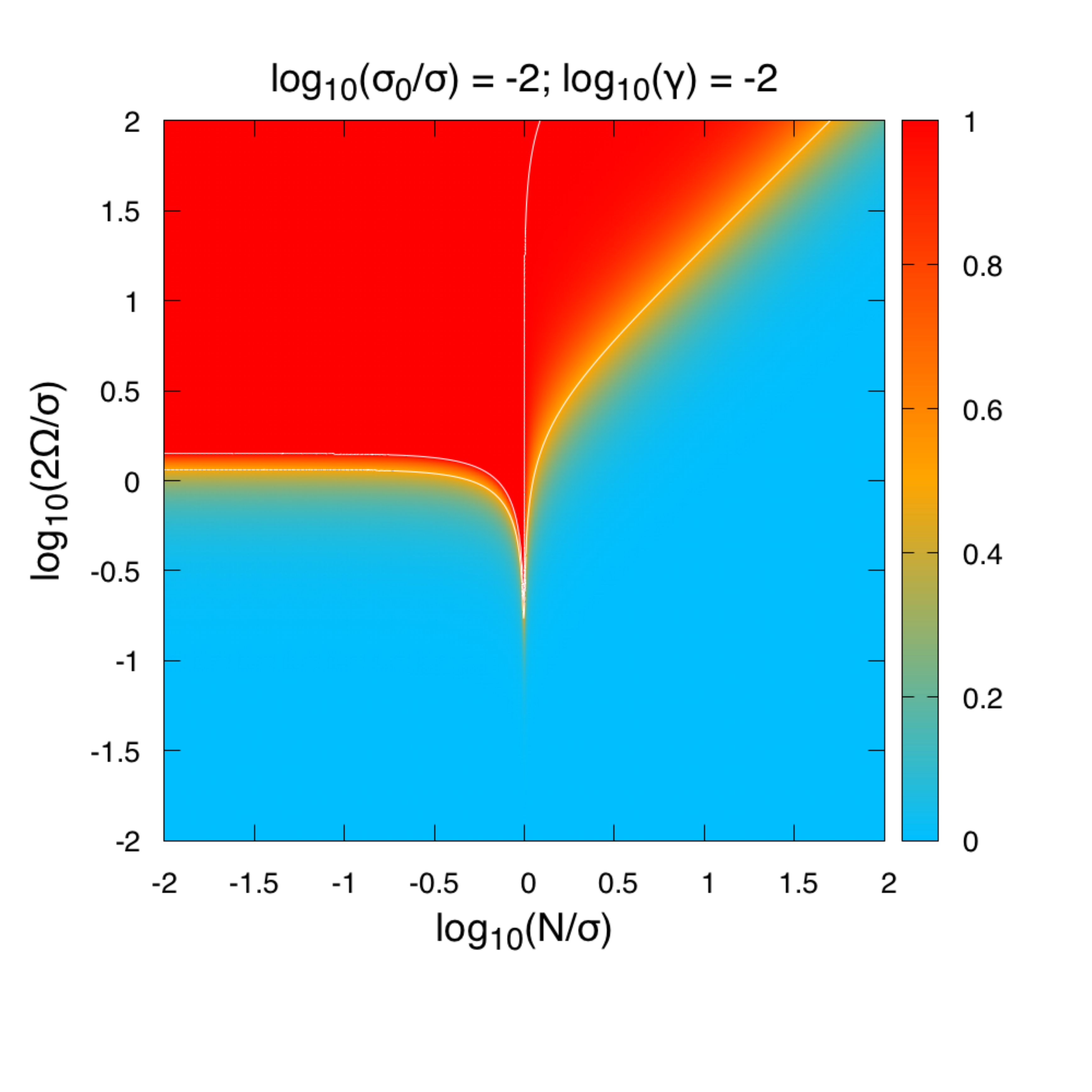}
   \includegraphics[width=0.31\textwidth,trim = 2.5cm 5.9cm 2.5cm 3.4cm, clip]{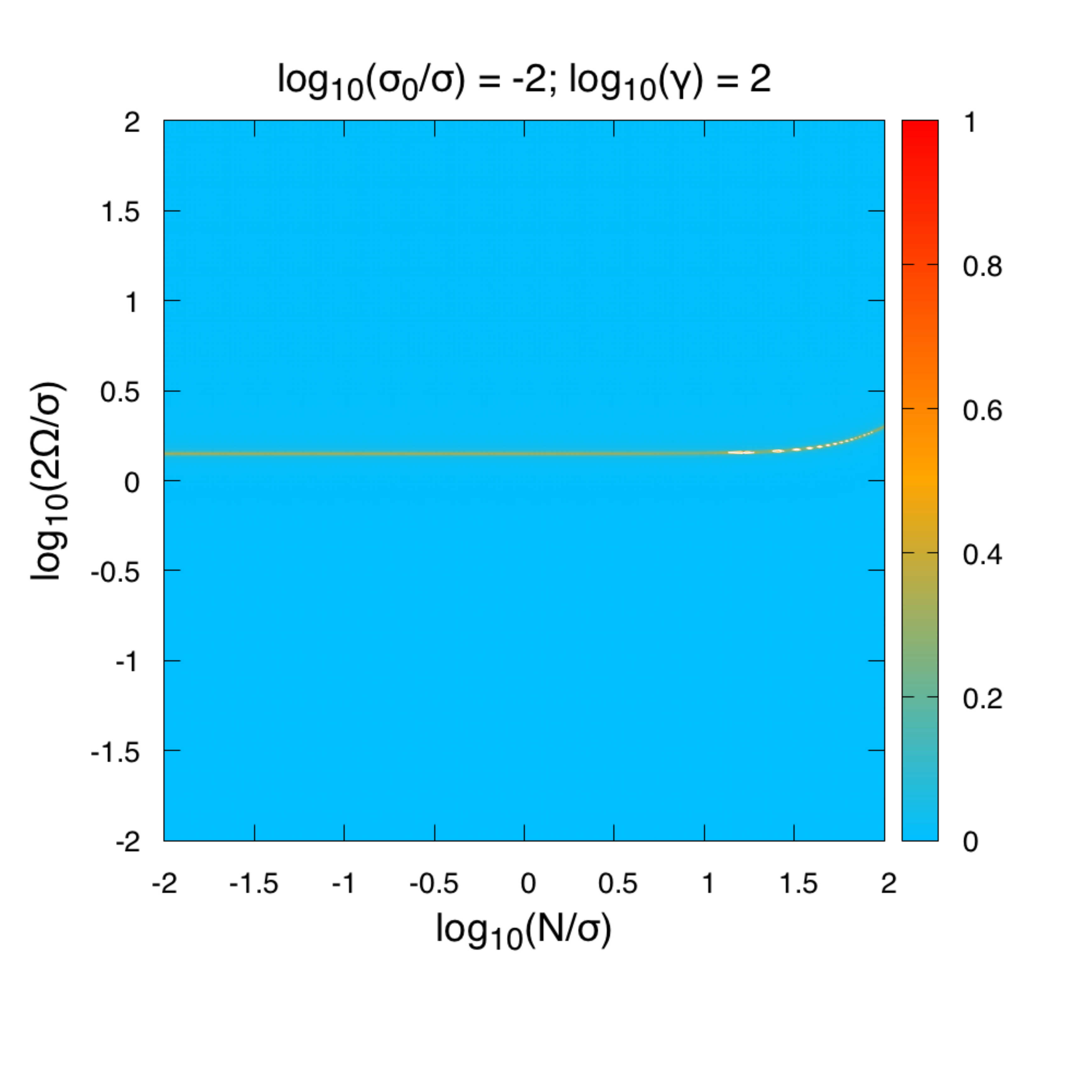} \\
   \hspace{1.0cm}
   \includegraphics[height=0.03\textwidth]{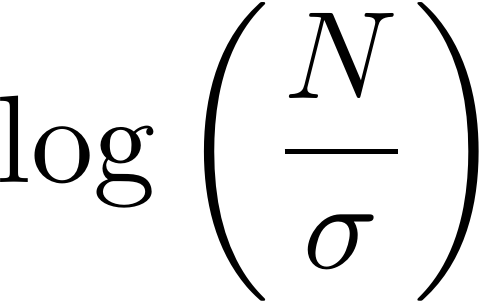} \hspace{5cm}
   \includegraphics[height=0.03\textwidth]{auclair-desrotour_fig5j.pdf}
   \caption{Relative difference between the parameters $ \nu_{\rm TA} $ and $ \nu_{\rm NTA} $ corresponding to the cases \emph{with} and \emph{without} traditional approximation respectively, as a function of the ratio $ N / \sigma $ (horizontal axis) and $ 2 \Omega / \sigma $ (vertical axis) in logarithmic scales, and for various values of the ratios $ \sigma_0 / \sigma $ and $ \gamma = \tau /  k $. From left to right, \smc{$ \log \left( \gamma \right) = \left\{ -2,2 \right\} $}. From bottom to top, \smc{$ \log \left( \sigma_0 / \sigma \right) = \left\{ -2,2 \right\} $}. The \smc{normalized} relative difference between $ \nu_{\rm TA} $ and $ \nu_{\rm NTA} $ is given by $ \varsigma  = \left| \nu_{\rm TA} - \nu_{\rm NTA} \right| / \left| \nu_{\rm TA} + \nu_{\rm NTA} \right| $. Blue (red) areas designate regions where the traditional approximation is (not) appropriate. Parameters: $ \theta = \pi / 4 $ and $ \alpha = \pi/2 $.}
\label{fig:map_ecart_nu}%
\end{figure*}

\section{Domain of validity of the traditional approximation}
\label{sec:app_trad}

As mentioned in Section~\ref{sec:dynamics}, the traditional approximation consists in ignoring terms \rec{involving the factor} $ \tilde{f} $ in the momentum equation \smc{components}, Eqs.~(\ref{eqprim1}-\ref{eqprim3}). This is convenient to eliminate the coupling between the horizontal and vertical structures of the tidal response induced by the Coriolis acceleration in global modelings \citep[e.g.][]{CL70,ADLM2016a}. However, the regime of parameters where the approximation is appropriate still remains only partially determined. Early studies generally agree on the fact that the approximation can be applied if $ 2 \Omega \ll \sigma $, that is in the regime of super-inertial waves, where the fluid tidal response is \smc{weakly} affected by the rotation of the planet. It has also been shown that the above condition could be extended to $ 2 \Omega \lesssim \sigma $ in the case of a stably-stratified fluid ($ \sigma \ll N $) \smc{\citep[e.g.][]{Friedlander1987,Mathis2008,Mathis2009,Prat2017}}. 

We propose here to quantify the conditions of \smc{applicability of} the traditional approximation by establishing the boundaries of its domain of validity as a function of the characteristic frequencies of the system (i.e. $ \sigma $, $ \sigma_0 $, $ 2 \Omega $, $ N $), and of the length scale ratio $ \gamma $. 

In light of the expressions of the dissipated power and tidal torque, \smc{given in} Eqs.~(\ref{Ddiss_iso}) and (\ref{torque_iso}), the dimensionless parameter $ \nu $ introduced in Eqs.~(\ref{nu_uni}) and (\ref{nu_iso}) appears as a key parameter to characterize the domain of validity of the traditional approximation. Indeed, it intervenes in an essential way in $ \hat{D}_{\rm diss} $ and $ \mathcal{T} $, and contains all the information concerning the hierarchy of characteristic frequencies and control parameters of the system. Therefore, considering that the impact of the traditional approximation on \smc{obtained} results is directly related to the variation of $ \nu $ with $ \tilde{f} $, we adopt as index of validity the \smc{normalized} relative difference between the parameters $ \nu_{\rm TA} $ and $ \nu_{\rm NTA} $ corresponding to the cases \emph{with} ($ \tilde{f} = 0 $) and \emph{without} ($ \tilde{f} \neq 0 $) \rec{the traditional approximation} respectively (the subscripts TA and NTA stand for \emph{Traditional Approximation} and \emph{No Traditional Approximation} \smc{respectively}), which is expressed as

\begin{equation}
\varsigma = \left| \frac{\nu_{\rm TA} - \nu_{\rm NTA}}{\nu_{\rm TA} + \nu_{\rm NTA} } \right|.
\end{equation}

\noindent In this approach, the condition of validity of the traditional approximation is $ \varsigma \ll 1 $. Otherwise, neglecting the latitudinal \smc{projection of the rotation vector} strongly modifies the tidal response\smc{, leading to $\varsigma \approx 1 $,} and the traditional approximation \smc{should be abandoned}. 

The parameter $ \varsigma $ is plotted \rec{in Fig.}~\ref{fig:map_ecart_nu} as a function of the frequencies ratios $ N/ \sigma $, $ 2 \Omega / \sigma $ and \smc{for asymptotic values of} $ \sigma_0 / \sigma $ and the length-scale parameter $ \gamma = \tau / k_\perp $. Angles are set to $ \theta = \pi / 12 $ and $ \alpha = \pi / 2 $. The values taken by $ \varsigma $ are indicated by colors. Blue regions indicate the domain of parameters where the traditional approximation \smc{can be applied} ($ \varsigma \ll 1 $) while red ones designate regimes where it is not relevant ($\varsigma \approx 1$). Hence, we note that the traditional approximation is appropriate whatever the hierarchy of characteristic frequencies if the wavelength of the mode is far greater than the pressure height-scale, i.e. if $ \gamma \gg 1 $. This is due to the fact that the term $ \left( 1 / 2 \right) \gamma^2 \left( \sigma^2 - f^2 \right) $ is always far greater in this case than terms associated with the latitudinal component of the Coriolis acceleration in the denominator of $ \nu $,

\begin{align}
\label{den}
d = & \ \eta N^2 - \sigma^2 + f_{\rm s}^2 + \gamma \tilde{f} \left( i f \sin \alpha - \sigma \cos \alpha \right) \\
 & + \frac{1}{2} \gamma^2 \left( \sigma^2 - f^2 \right) . \nonumber
\end{align}

This suggests that the traditional approximation is well adapted to the treatment of the tidal response of thin atmospheres, where the horizontal wavelength of dominating propagating modes is comparable to the radius of the planet in order of magnitude. In the case of small wavelengths ($ \gamma \ll 1 $), we recover the domain of validity established by early studies\smc{:} $ 2 \Omega \ll \sigma $ if the layer is weakly stably-stratified or convective ($ N \ll \sigma $), and $ 2 \Omega \lesssim \sigma $ \smc{in the case of strong} stable stratification ($ N \gg \sigma $). Particularly, the \smc{bottom} left panel of Fig.~\ref{fig:map_ecart_nu} show that the boundary of the domain of validity of the approximation corresponds to $ 2 \Omega \sim N $ in the regime of strong stable stratification, \smc{which is in good agreement with the diagnosis done in \cite{ADLM2016a} (Fig.~21)}. This boundary is slightly modified by the radiative/diffusive cooling (top left panel), with a downward translation of magnitude $ \left( 1/2 \right) \log \left( \sigma / \sigma_0 \right) $ resulting from the \smc{equality} of the dominating terms of Eq.~(\ref{den})\smc{:} $ \left( \sigma / \sigma_0 \right) N^2 = f_{\rm s}^2 $.

\section{Conclusions}
\label{sec:conclusions}

Motivated by the understanding of the role played by thermally forced gravity waves in the atmospheric tidal response of a terrestrial planet \smc{submitted to the irradiation of its host star}, we have studied the tidal perturbation of a local \smc{atmospheric} section in solid rotation with the body. In this \emph{ab initio} approach, \rec{inspired by} \cite{GS2005} and \cite{MNT2014}, the dynamics of tidal waves are reduced to the essentials, which allowed us to conserve the whole physics of tides, and particularly \smc{all the components of the Coriolis acceleration while} avoiding \smc{complexities} associated with spherical geometry. The goal of this work was to provide a diagnosis of the results obtained in \smc{our} previous \smc{global analytic} study \citep[][]{ADLM2016a} \smc{where two asymptotic regimes were identified: a weakly stratified regime inducing a Maxwell-like tidal torque \smc{applied on the atmosphere} and a strongly stably-stratified one associated with a weak torque}.

We wrote the linearized equations of tides for a fluid section characterized by radial background distributions and dissipative processes \smc{modelled} with a Newtonian cooling as in \cite{ADLM2016a}. We computed analytic solutions describing the forced oscillatory response in two typical cases: a homogeneous fluid with uniform background distributions and an isothermal atmosphere. In both cases, the tidal torque exerted and the energy dissipated by diffusive/radiative processes were also derived analytically. We  showed that tidal regimes are defined by a small number of physical parameters, namely the tidal ($ \sigma $), inertia ($ 2 \Omega $), Brunt-Väisälä ($ N $) and radiative ($\sigma_0$) frequencies, and the ratio between the horizontal wavelength of a propagating \smc{tidal wave} and the pressure height-scale of the fluid ($\gamma $).  

Considering the semidiurnal stellar tide, we recovered the results obtained with global modelings in early works. \smc{The} dependence of the tidal torque on the forcing frequency can be described by a Maxwell model in the vicinity of spin-orbit synchronous rotation \citep[][]{Leconte2015,ADLM2016a} \smc{in the case of} weakly stably-stratified or convective atmospheres. If the stratification is strong, gravito-inertial waves are generated. This wavelike response breaks the \smc{large} scale quadrupolar hydrostatic elongation caused by the stellar heating at zero order, so that there is no net tidal bulge. As a consequence, the amplitude of the tidal torque diminishes and becomes negligible compared to the convective \smc{case}. This suggests that Venus-like planets with a stably-stratified atmosphere are likely to tend towards spin-orbit synchronous rotation while those with a convective atmosphere rather evolve towards non-synchronized states like Venus, as predicted by the model of \cite{ADLM2016a}. \smc{Hence}, the Cartesian approach \smc{allows us to explore the domain of parameters, to confirm the previously identified asymptotic regimes and to highlight the continuous transition between them}. \smc{This transition} occurs for $ N \sim \left\{ \gamma \sigma , \gamma \sigma_0 \right\} $. \rec{In the case of Venus itself, the structure of the atmosphere is characterized by a strong \rec{negative} temperature gradient near the surface of the planet \citep[][]{Seiff1980}. This implies that the densest layers of the atmosphere, which are also the better viscously coupled to the solid part, are neutrally stratified. Hence, our simplified modeling suggests in this case a strong atmospheric torque that agrees with the conclusions of early works dealing the rotation of Venus \citep[e.g.][]{GS1969,DI1980,CL01,ADLM2016b}.}

By isolating in \smc{obtained} solutions a characteristic scaling parameter ($ \nu $), we were \smc{also} able to delimit the domain of validity of the traditional approximation \smc{used in global models} as a function of the system's control parameters, and to characterize \smc{the} asymptotic regimes. In the short wavelengths approximation, we recovered the domain of validity identified in early studies, namely the validity conditions $ 2 \Omega \ll \sigma $ in the case of weakly stratified fluids, and $ 2 \Omega \ll N $ in the case of strongly stratified ones \smc{\citep[][]{Friedlander1987,Mathis2008,Mathis2009,ADLM2016a}}. This domain is slightly modified by dissipative mechanisms in this case. In the \rec{long wavelength asymptotic regime}, where the pressure \rec{scale height} is smaller than the \smc{horizontal} wavelength of a propagating mode, we found that the Coriolis terms neglected in the traditional approximation do not affect the tidal response, except at the boundary of the inertial regime, defined by $ 2 \Omega \sim \sigma $. This result is appropriate to characterize the forced response of the fluid up to meso-scale modes, that is the application limit of the f-plane approximation used \smc{in} our modeling. Therefore, the impact of the traditional approximation on global tidal modes \rec{should be studied} \smc{in general} with global modelings, as done by \cite{OL2004} in the case of giant planets or \cite{Tort2014} for atmospheric dynamics.

However, the local approach developed here still remains an efficient way to peer into the large domain of parameters of \smc{atmospheric} tidal responses and to clarify the predictions of global modelings \smc{obtained with assumptions}. \smc{Indeed,} it offers a simplified \smc{but robust} picture of the tidal waves dynamics \smc{with the essential physical ingredients}. \smc{Its predictive power on the rotation state of exoplanets as a function of the convective stability of their atmosphere is of great interest with their probe by forthcoming space missions like JWST \citep[][]{Lagage2015}. Such missions will be able to provide constraints on the temperature gradient at different depths in the atmospheres of exoplanets thanks to multi-wavelenght observations in the near IR.} In forthcoming studies, we will complete this picture by including in the Cartesian fluid section stratified background flows. \smc{These flows} are likely to strongly modify the atmospheric tidal response but cannot be taken into account \smc{directly in global analytic modelings} without important mathematical complications.

\begin{acknowledgements}
      P. Auclair-Desrotour and S. Mathis acknowledge funding by the European Research Council through ERC grants WHIPLASH 679030 and SPIRE 647383. \rec{The authors wish to thank the anonymous referee for helpful suggestions and remarks.}
\end{acknowledgements}

\bibliographystyle{aa} 
\bibliography{references} 

\end{document}